# A Generalized Converted Measurement Kalman Filter

**Steven V. Bordonaro**
**Undersea Warfare Platforms and Payload Integration Department**

**Tod E. Luginbuhl**
**Michael J. Walsh**
**Sensors and Sonar Systems Department**

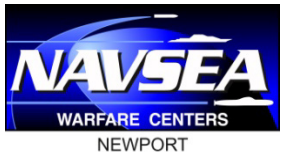

# Naval Undersea Warfare Center Division Newport, Rhode Island

# PREFACE

This report was prepared under Naval Undersea Warfare Center Division Newport (NUWCDIVNPT) project, "Derivation of a Generalized Kalman Filter for Nonlinear Measurements with Improved Stability Over Kalman Filter Extensions," principal investigator Steven V. Bordonaro (Code 454). The report was funded by a NUWCDIVNPT ILIR project.

The technical reviewer for this report was Mark P. Lowney (Code 8511).

Reviewed and Approved: 18 July 2024

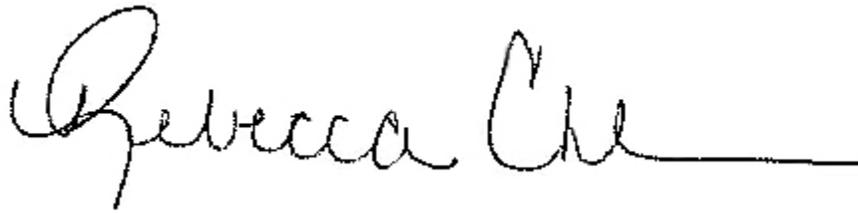

**Rebecca Chhim**
**Head (Acting), Undersea Warfare Platforms and Payload Integration Department**

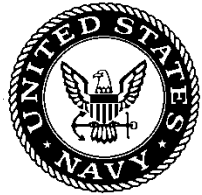

# REPORT DOCUMENTATION PAGE

| 1. REPORT DATE | 2. REPORT TYPE | 3. DATES COVERED | |
|---|---|---|---|
| 18-10-2023 | Technical Report | START DATE | END DATE |

**4. TITLE AND SUBTITLE**

A Generalized Converted Measurement Kalman Filter

| 5a. CONTRACT NUMBER | 5b. GRANT NUMBER | 5c. PROGRAM ELEMENT NUMBER |
|---|---|---|
| | | |
| **5d. PROJECT NUMBER** | **5e. TASK NUMBER** | **5f. WORK UNIT NUMBER** |
| | | |

**6. AUTHOR(S)**

Steven V. Bordonaro, Tod E. Luginbuhl, Michael J. Walsh

| 7. PERFORMING ORGANIZATION NAME(S) AND ADDRESS(ES) | 8. PERFORMING ORGANIZATION REPORT NUMBER |
|---|---|
| Naval Undersea Warfare Center Division<br>1176 Howell Street<br>Newport, RI 02841-1708 | TR 12,472 |

| 9. SPONSORING/MONITORING AGENCY NAME(S) AND ADDRESS(ES) | 10. SPONSOR/MONITOR'S ACRONYM(S) | 11. SPONSOR/MONITOR'S REPORT NUMBER(S) |
|---|---|---|
| Naval Undersea Warfare Center Division<br>1176 Howell Street<br>Newport, RI 02841-1708 | NUWCDIVNPT | TR 12,472 |

**12. DISTRIBUTION/AVAILABILITY STATEMENT**

DISTRIBUTION STATEMENT A. Approved for public release; distribution is unlimited.

**13. SUPPLEMENTARY NOTES**

**14. ABSTRACT**

This report derives a generalized, converted measurement Kalman filter for the class of filtering problems with a linear state equation and nonlinear measurement equation, for which a bijective mapping exists between the state and measurement coordinate systems. For these problems, a procedure is developed for mapping the observed measurements and their covariance matrices from measurement coordinates to state coordinates, such that the converted measurements are unbiased and the converted measurement covariance matrices are independent of the states and observed measurements. In cases where not all measurement coordinates are observed, predicted measurements of these coordinates are introduced as substitutes, and the impact of these measurements on the filter is mitigated by an information zeroing operation on the corresponding rows and columns of the converted measurement inverse-covariance matrix. Filter performance is demonstrated on two well-known target-tracking problems and is compared with the performance of the standard extended and unscented Kalman filters for these problems. These examples show the proposed filter obtains lower mean squared error, better consistency, and less track loss than either the extended Kalman filter or the unscented Kalman filter.

# TABLE OF CONTENTS

# LIST OF ILLUSTRATIONS

## LIST OF ABBREVIATIONS AND ACRONYMS

| | |
|---|---|
| ANEES | Average Normalized Estimation Error Squared |
| CI | Confidence Interval |
| CKF | Cubature Kalman Filter |
| EKF | Extended Kalman Filter |
| MSE | Mean Squared Error |
| PDF | Probability Density Function |
| PKF | Precision Kalman Filter |
| PCRLB | Posterior Cramér-Rao Lower Bound |
| SPT | Sigma Point Transform |
| UKF | Unscented Kalman Filter |

# 1. INTRODUCTION

In many familiar filtering applications, it is often appropriate to model the problem using two distinct coordinate systems: one natural to the state process, in which the state dynamics are linear, and one natural to the measurement process, where the mapping between the two coordinate systems is nonlinear. For example, in target tracking applications, it is often natural and convenient to model the target state (position, velocity, etc.) in Cartesian coordinates, while the physics and geometry of the sensing system typically dictate curvilinear coordinates (e.g., polar, spherical, cylindrical, etc.) as the natural coordinates. Such problems require generalizations of the standard Kalman filter—such as the extended Kalman filter (EKF) (references 1 and 2), or a Gaussian filter (reference 3) such as the unscented Kalman filter (UKF) (references 4–6)—to deal with the nonlinear mapping between the state and measurement coordinate systems.

An alternative approach is to map the measurements (and their covariance matrices) from measurement coordinates to state coordinates and apply the standard (linear) Kalman filter to the converted measurements. Since the nonlinear mapping from measurement to state coordinates is likely to yield converted measurements with non-Gaussian statistics, care must be taken to design a procedure that produces converted measurements that best approximate the standard Kalman filtering assumptions. As stated in the abstract of reference 7:

> "There are two potential issues that arise when performing converted measurement tracking. The first is conversion bias that occurs when the measurement transformation introduces a bias in the expected value of the converted measurement. The second is estimation bias that occurs because the estimation of the converted measurement error covariance is correlated with the measurement noise, leading to a biased Kalman gain."

References 7–10 develop a measurement conversion procedure for an increasingly complex sequence of standard, practical target tracking problems that address the first issue and mitigate the second. This procedure produces converted measurements that are unbiased, and converted measurement covariance matrices that are independent of the state estimates and observed measurements, resulting in consistent linear Kalman filters that outperform the EKF and UKF. This measurement conversion procedure is generalized in this report to the class of filtering problems with a linear state equation, and nonlinear measurement equation for which a bijective mapping exists between the state and measurement coordinate systems.

# 2. NOTATIONS AND CONVENTIONS

Let $\mathcal{X}$ and $\mathcal{Z}$ denote the state and measurement coordinate systems, respectively, each a subset of $\mathbb{R}^N$, and let $h : \mathcal{X} \to \mathcal{Z}$ and $g : \mathcal{Z} \to \mathcal{X}$ denote the mappings between these coordinate systems. Let $\mathbf{x}(k)$ and $\mathbf{z}(k)$ denote the state and measurement vectors at time index, $k$, where $k = 0, 1, \ldots$. The derivations in this report involve mapping state and measurement vectors and related quantities back and forth between the state and measurement coordinates systems. To clarify which coordinate system is being referenced, $\mathcal{X}$ and $\mathcal{Z}$ are used as subscripts. Hence, $\mathbf{x}_\mathcal{X}$ refers to the state in state coordinates, whereas $\mathbf{x}_\mathcal{Z} = h(\mathbf{x}_\mathcal{X})$ refers to the state in measurement coordinates. Likewise, $\mathbf{z}_\mathcal{Z}$ refers to the measurement in measurement coordinates, whereas $\mathbf{z}_\mathcal{X} = g(\mathbf{z}_\mathcal{Z})$ refers to the measurement in state coordinates.

The overset symbols, "ˆ" and "˜", are used in this report to denote estimates and errors, respectively, while the symbol "|" is used to denote probabilistic conditioning. For example, $\hat{\mathbf{x}}_\mathcal{X}(k|k)$ denotes the estimate of the state, in state coordinates, at time index, $k$, given all measurements up through $k$, and

$$\tilde{\mathbf{x}}_\mathcal{X}(k|k) = \hat{\mathbf{x}}_\mathcal{X}(k|k) - \mathbf{x}_\mathcal{X}(k) \tag{2-1}$$

denotes the error in this estimate.

The derivations in this report require the Jacobian matrices of the mappings, $h$ and $g$. Let $\mathrm{J}_h : \mathcal{X} \to \mathbb{R}^{N \times N}$ and $\mathrm{J}_g : \mathcal{Z} \to \mathbb{R}^{N \times N}$ denote the Jacobain matrices of $h$ and $g$, respectively. Specifically, let $\nabla_\mathbf{x}$ denote the vector partial derivative operator with respect to $\mathbf{x} = \begin{bmatrix} x_1 & x_2 & \ldots & x_N \end{bmatrix}^t$,

$$\nabla_\mathbf{x} = \begin{bmatrix} \frac{\partial}{\partial x_1} & \frac{\partial}{\partial x_2} & \cdots & \frac{\partial}{\partial x_N} \end{bmatrix}^t, \tag{2-2}$$

where the superscript, $t$, denotes vector-matrix transpose. Then, for any $\mathbf{x}_* \in \mathcal{X}$, the Jacobian matrix of $h$, evaluated at $\mathbf{x}_*$, is given by

$$\mathrm{J}_h(\mathbf{x}_*) = \left[\nabla_\mathbf{x} h^t(\mathbf{x}_*)\right]^t. \tag{2-3}$$

The Jacobian matrix of $g$ is defined similarly.

Finally, not all of the measurement coordinates may be physically measured (or observed). Let $\mathcal{Z}_m \subseteq \mathcal{Z}$ denote the $M$-dimensional subset of $\mathcal{Z}$ that is measured with $M \leq N$, and let $\mathcal{Z}_u = \mathcal{Z} \setminus \mathcal{Z}_m$ denote the subset of $\mathcal{Z}$ that is unmeasured (or unobserved). Then the measurement vector, $\mathbf{z}_\mathcal{Z}$, may be partitioned as

$$\mathbf{z}_\mathcal{Z} = \begin{bmatrix} \mathbf{z}_{\mathcal{Z}_m} \\ \mathbf{z}_{\mathcal{Z}_u} \end{bmatrix}, \tag{2-4}$$

with $\mathbf{z}_{\mathcal{Z}_m} \in \mathcal{Z}_m$ and $\mathbf{z}_{\mathcal{Z}_u} \in \mathcal{Z}_u$. Likewise, the mapping, $h$, from state coordinates to measurement coordinates may be partitioned as $h = \begin{bmatrix} h_m^t & h_u^t \end{bmatrix}^t$, where $h_m : \mathcal{X} \to \mathcal{Z}_m$ and $h_u : \mathcal{X} \to \mathcal{Z}_u$ are the mappings from the state coordinate system to the observed and unobserved measurement coordinate systems, respectively. If $\mathrm{R}_{\mathcal{Z}_m}$ and $\mathrm{R}_{\mathcal{Z}_u}$ denote the covariance matrices associated with $\mathbf{z}_{\mathcal{Z}_m}$ and $\mathbf{z}_{\mathcal{Z}_u}$, respectively, and $\mathrm{R}_{\mathcal{Z}_m \mathcal{Z}_u}$ denotes the cross-covariance matrix, then the covariance matrix associated with the measurement vector, $\mathbf{z}_\mathcal{Z}$, is partitioned as

$$\mathrm{R}_\mathcal{Z} = \begin{bmatrix} \mathrm{R}_{\mathcal{Z}_m} & \mathrm{R}_{\mathcal{Z}_m \mathcal{Z}_u} \\ \mathrm{R}_{\mathcal{Z}_m \mathcal{Z}_u}^t & \mathrm{R}_{\mathcal{Z}_u} \end{bmatrix}. \tag{2-5}$$

Procedures for specifying quantitative values for the unobserved terms, $\mathbf{z}_{\mathcal{Z}_u}$, $\mathrm{R}_{\mathcal{Z}_u}$, and $\mathrm{R}_{\mathcal{Z}_m \mathcal{Z}_u}$, for practical applications of the filter developed in this report are discussed in the following sections.

# 3. PROBLEM STATEMENT

The nonlinear filtering problem of interest in this report has a linear state equation and nonlinear measurement equation which may be expressed as follows:

$$\mathbf{x}_{\mathcal{X}}(k) = \mathrm{A}(k)\,\mathbf{x}_{\mathcal{X}}(k-1) + \tilde{\mathbf{q}}_{\mathcal{X}}(k), \tag{3-1}$$

$$\mathbf{z}_{\mathcal{Z}_m}(k) = h_m(\mathbf{x}_{\mathcal{X}}(k)) + \tilde{\mathbf{z}}_{\mathcal{Z}_m}(k), \tag{3-2}$$

where $\mathrm{A}(k)$ is a known, $N \times N$ transition matrix; $\tilde{\mathbf{q}}_{\mathcal{X}}(k)$ and $\tilde{\mathbf{z}}_{\mathcal{Z}_m}(k)$ are independent, Gaussian distributed process and measurement noise vectors with zero means and known covariance matrices, $\mathrm{Q}_{\mathcal{X}}(k)$ and $\mathrm{R}_{\mathcal{Z}_m}(k)$, respectively, and $\tilde{\mathbf{q}}_{\mathcal{X}}(j)$ and $\tilde{\mathbf{z}}_{\mathcal{Z}_m}(k)$ are independent of one another for all $j$ and $k$. The objective of this filtering problem is to find the state estimate, $\hat{\mathbf{x}}_{\mathcal{X}}(k|k)$, at each time index, $k$, that minimizes the mean squared error (MSE),

$$\mathrm{E}\left\{ [\hat{\mathbf{x}}_{\mathcal{X}}(k|k) - \mathbf{x}_{\mathcal{X}}(k)]^t\,[\hat{\mathbf{x}}_{\mathcal{X}}(k|k) - \mathbf{x}_{\mathcal{X}}(k)] \right\}, \tag{3-3}$$

given all measurements up through $k$, that is, $\{\mathbf{z}_{\mathcal{Z}_m}(1), \mathbf{z}_{\mathcal{Z}_m}(2), \ldots, \mathbf{z}_{\mathcal{Z}_m}(k)\}$. The problem is complicated by the nonlinear mapping, $h_m$, from state to measurement coordinates, and its general solution within the Kalman filtering framework requires approximations such as those used by the EKF or the Gaussian filter (e.g., the UKF).

Alternatively, many researchers, particularly in target tracking, have sought to turn this nonlinear filtering problem, for specific applications, into a linear filtering problem by directly mapping the observed measurements, $\mathbf{z}_{\mathcal{Z}_m}$, from the observed measurement coordinate system, $\mathcal{Z}_m$, to the appropriate subset of the state coordinate system, $\mathcal{X}$. For example, in references 8, 11–16 several researchers developed "converted measurement" linear Kalman filters for the problem of tracking a constant velocity target in Cartesian coordinates by converting the measured polar coordinates (target range and bearing) to position in Cartesian coordinates. The obvious difficulty with these approaches lies in accounting for the fact the distributions of the converted measurements are almost certainly not Gaussian due to the nonlinear mapping from $\mathcal{Z}_m$ to $\mathcal{X}$ (see the discussion in reference 17), which violates an assumption of the standard Kalman filter. This difficulty often results in converted measurement equations and filter recursion that are functions of the true (unknown) values of the states, or approximations to these values using the measurements themselves, which introduce biases and inconsistencies in the state estimates and their covariances, respectively.

Reference 7 classifies these problems in terms of conversion and estimation biases. The first type of bias is caused by the measurement transformation, $h$, and the second occurs when the converted measurement covariance matrix is correlated with the measurement noise. References 7–10 develop a measurement conversion procedure, for an increasingly complex sequence of target tracking problems, that address these two issues and yield consistent state estimates using the information form of the linear Kalman filter recursions. This report generalizes this procedure to the class of filtering problems with a linear state equation and nonlinear measurement equation, for which a bijective mapping exists between the state and measurement coordinates. However, there is no guarantee the resulting state estimates minimize the MSE in equation (3-3) because the procedure requires the converted measurement to be unbiased. Reference 8 shows this is true when the measurements are in polar and spherical coordinates and the target state vectors are in Cartesian coordinates.

For the special case when all of the measurement coordinates are observed, that is, when $\mathcal{Z}_u = \emptyset$, this generalized procedure replaces the nonlinear measurement equation,

$$\mathbf{z}_z(k) = h(\mathbf{x}_x(k)) + \tilde{\mathbf{z}}_z(k), \tag{3-4}$$

with the linear measurement equation,

$$\bar{\mathbf{z}}_x(k) = \mathbf{x}_x(k) + \tilde{\mathbf{z}}_x(k), \tag{3-5}$$

where $\bar{\mathbf{z}}_x(k)$ is a debiased version of the raw converted measurement, $\mathbf{z}_x(k) = g(\mathbf{z}_z(k))$, and $\tilde{\mathbf{z}}_x(k)$ is a Gaussian distributed measurement noise vector, in state coordinates, with zero mean and known covariance matrix $\bar{\mathrm{R}}_x(k)$. General expressions for $\bar{\mathbf{z}}_x(k)$ and $\bar{\mathrm{R}}_x(k)$ are derived in sections 6 and 7, respectively. For this case, the state estimate, $\hat{\mathbf{x}}_x(k|k)$, is obtained from the linear Kalman filter recursions resulting from the state equation, equation (3-1), and the converted measurement equation, equation (3-5).

For the general case when only some of the measurement coordinates are observed (i.e., $\mathcal{Z}_u \neq \emptyset$), the predicted measurements at time index, $k$, obtained from the predicted state estimate, $\hat{\mathbf{x}}_x(k|k-1)$, are substituted for the unobserved measurements, $\mathbf{z}_{z_u}(k)$, as discussed in section 6. To compensate for this addition of information, these predicted measurements are largely rendered non-informative by an "information zeroing" operation, developed in section 7. The "information zeroing" operation zeros out the rows and columns corresponding to the unobserved measurements in the inverse covariance matrix, $\bar{\mathrm{R}}_x^{-1}(k)$, reducing the rank of this matrix by $N - M$. Since the converted measurement inverse-covariance (or precision) matrix, $\bar{\mathrm{R}}_x^{-1}(k)$, is not invertible in this case, the standard Kalman filter recursions are replaced with their information forms (see section 7.2 of reference 1) as shown in section 5. The new generalized, converted measurement Kalman filter described in this report is called the precision Kalman filter (PKF) because the converted measurement precision matrix, $\bar{\mathrm{R}}_x^{-1}(k)$, is central to the filter's effectiveness.

## 4. FUNDAMENTAL ASSUMPTIONS

The derivations of the converted measurement debiasing function and precision matrix in the following sections rely on a few fundamental assumptions, which are stated explicitly below:

**Assumption 1.** The nonlinear mappings between the state and measurement coordinate systems are measurable, integrable, and bijective. The inverse mappings, $h^{-1}$ and $g^{-1}$, exist, and are related by $h^{-1} = g$ and $g^{-1} = h$.

**Assumption 2.** The mappings $h$ and $g$ are continuously differentiable almost everywhere in $\mathcal{X}$ and $\mathcal{Z}$, respectively. This assumption, together with assumption 1 and the Inverse Function Theorem, imply that if $\mathrm{J}_h(\mathbf{x}_x)$ is invertible at $\mathbf{x}_x = g(\mathbf{x}_z) \in \mathcal{X}$ (that is, if $|\mathrm{J}_h(\mathbf{x}_x)| \neq 0$), then

$$\mathrm{J}_g(\mathbf{x}_z) = \mathrm{J}_h^{-1}(\mathbf{x}_x). \tag{4-1}$$

Likewise, if $J_g(\mathbf{x}_z)$ is invertible at $\mathbf{x}_z = h(\mathbf{x}_x) \in \mathcal{Z}$ (that is, if $|J_g(\mathbf{x}_z)| \neq 0$), then

$$J_h(\mathbf{x}_x) = J_g^{-1}(\mathbf{x}_z). \tag{4-2}$$

**Assumption 3.** When only some of the measurement coordinates are observed (i.e., $\mathcal{Z}_u \neq \emptyset$), predicted measurements at time index, $k$, based on the predicted state estimate, $\hat{\mathbf{x}}_x(k|k-1)$, may be substituted for the unobserved measurements, $\mathbf{z}_{\mathcal{Z}_u}(k)$:

$$\mathbf{z}_{\mathcal{Z}_u}(k) = h_u(\hat{\mathbf{x}}_x(k|k-1)). \tag{4-3}$$

The contribution of these predicted measurements to the state estimate, $\hat{\mathbf{x}}_x(k)$, is rendered largely non-informative by the information zeroing operation developed in section 7.

**Assumption 4.** The complete measurement vector, $\mathbf{z}_z(k) = \begin{bmatrix} \mathbf{z}_{\mathcal{Z}_m}^t(k) & \mathbf{z}_{\mathcal{Z}_u}^t(k) \end{bmatrix}^t$, which is the concatenation of the observed and unobserved measurements, obeys equation (3-4), where $\tilde{\mathbf{z}}_z(k)$ is a Gaussian distributed noise vector with zero mean and known covariance matrix, $R_z(k)$, given by equation (2-5). Quantitative values for the covariance matrix, $R_{z_u}(k)$, and cross-covariance matrix, $R_{z_m z_u}(k)$, are determined, in practice, from problem-specific prior information.

**Assumption 5.** The true (unknown) value of the state at time index, $k$, in measurement coordinates, $\mathbf{x}_z(k) = h(\mathbf{x}_x(k))$, is approximately equal to the first order Taylor expansion about the predicted state, $\hat{\mathbf{x}}_x(k|k-1)$:

$$\mathbf{x}_z(k) \approx h(\hat{\mathbf{x}}_x(k|k-1)) - \tilde{\mathbf{x}}_z(k|k-1), \tag{4-4}$$

where the predicted state error in measurement coordinates, $\tilde{\mathbf{x}}_z(k|k-1)$, is defined as

$$\tilde{\mathbf{x}}_z(k|k-1) = J_h(\hat{\mathbf{x}}_x(k|k-1))\, \tilde{\mathbf{x}}_x(k|k-1). \tag{4-5}$$

The first order Taylor expansion given in equation (4-4) is derived as follows. First the state equation, equation (3-1), is written in terms of the state estimate and its error at time index, $k-1$:

$$\begin{align}
\mathbf{x}_x(k) &= A(k)\, \mathbf{x}_x(k-1) + \tilde{\mathbf{q}}_x(k), \tag{4-6}\\
&= A(k)\, [\hat{\mathbf{x}}_x(k-1|k-1) - \tilde{\mathbf{x}}_x(k-1|k-1)] + \tilde{\mathbf{q}}_x(k), \tag{4-7}\\
&= A(k)\, \hat{\mathbf{x}}_x(k-1|k-1) - [A(k)\, \tilde{\mathbf{x}}_x(k-1|k-1) - \tilde{\mathbf{q}}_x(k)], \tag{4-8}\\
&= \hat{\mathbf{x}}_x(k|k-1) - \tilde{\mathbf{x}}_x(k|k-1), \tag{4-9}
\end{align}$$

where the predicted state, $\hat{\mathbf{x}}_x(k|k-1)$, and its error, $\tilde{\mathbf{x}}_x(k|k-1)$, are defined by the first and second terms of the difference in equation (4-8), respectively.

$$\begin{align}
\mathbf{x}_z(k) &= h(\mathbf{x}_x(k)), \tag{4-10}\\
&= h(\hat{\mathbf{x}}_x(k|k-1) - \tilde{\mathbf{x}}_x(k|k-1)), \tag{4-11}\\
&\approx h(\hat{\mathbf{x}}_x(k|k-1)) - J_h(\hat{\mathbf{x}}_x(k|k-1))\, \tilde{\mathbf{x}}_x(k|k-1). \tag{4-12}
\end{align}$$

The approximation in equation (4-12) is the first-order Taylor series expansion of $h$ about the predicted state, $\hat{\mathbf{x}}_x(k|k-1)$, a byproduct of the PKF recursions given in section 5. The approximation given by equation (4-4) follows from equation (4-12), by defining the predicted state error, in measurement coordinates, as in equation (4-5).

**Assumption 6.** The estimated state error, $\tilde{\mathbf{x}}_x(k) = \hat{\mathbf{x}}_x(k) - \mathbf{x}_x(k)$, is Gaussian distributed with zero mean and covariance matrix, $P_x(k|k)$. From equation (4-8), this implies the predicted state error (in state coordinates), $\tilde{\mathbf{x}}_x(k|k-1)$, is Gaussian distributed with zero mean and covariance matrix

$$P_x(k|k-1) = A(k) P_x(k-1|k-1) A^t(k) + Q_x(k). \tag{4-13}$$

Moreover, from equations (4-4) and (4-5), this implies the predicted state error in measurement coordinates, $\tilde{\mathbf{x}}_z(k|k-1)$, is approximately Gaussian distributed with zero mean and covariance matrix

$$P_z(k|k-1) \approx J_h(\hat{\mathbf{x}}_x(k|k-1)) P_x(k|k-1) J_h^t(\hat{\mathbf{x}}_x(k|k-1)). \tag{4-14}$$

In the sequel, it will be useful to define the combined error, $\tilde{\mathbf{v}}_z(k)$, as the difference between the predicted state error, $\tilde{\mathbf{x}}_z(k|k-1)$, and the measurement error, $\tilde{\mathbf{z}}_z(k)$,

$$\tilde{\mathbf{v}}_z(k) = \tilde{\mathbf{x}}_z(k|k-1) - \tilde{\mathbf{z}}_z(k). \tag{4-15}$$

It follows from this assumption and assumption 4, that $\tilde{\mathbf{v}}_z(k)$ is approximately Gaussian distributed with zero mean and covariance matrix, $P_z(k|k-1) + R_z(k)$.

**Assumption 7.** Let $\mathbf{u}$ be random vector with known mean, $\boldsymbol{\mu}_\mathbf{u}$, and covariance matrix, $C_\mathbf{u}$, and let $\mathbf{w} = f(\mathbf{u})$, for some differentiable and integrable function, $f$. Then, to a first order approximation, $\mathbf{w}$ has mean $\boldsymbol{\mu}_\mathbf{w} = f(\boldsymbol{\mu}_\mathbf{u})$, and covariance matrix, $C_\mathbf{w} = J_f(\boldsymbol{\mu}_\mathbf{u}) C_\mathbf{u} J_f^t(\boldsymbol{\mu}_\mathbf{u})$. Indeed, since, to a first order approximation,

$$\begin{aligned}
\mathbf{w} &= f(\mathbf{u}), \\
&= f(\boldsymbol{\mu}_\mathbf{u} + (\mathbf{u} - \boldsymbol{\mu}_\mathbf{u})), \\
&\approx f(\boldsymbol{\mu}_\mathbf{u}) + J_f(\boldsymbol{\mu}_\mathbf{u})(\mathbf{u} - \boldsymbol{\mu}_\mathbf{u}),
\end{aligned} \tag{4-16}$$

it follows that $\boldsymbol{\mu}_\mathbf{w} = E\{\mathbf{w}\} \approx f(\boldsymbol{\mu}_\mathbf{u})$, and

$$\begin{aligned}
C_\mathbf{w} &= E\left\{(\mathbf{w} - \boldsymbol{\mu}_\mathbf{w})(\mathbf{w} - \boldsymbol{\mu}_\mathbf{w})^t\right\}, \\
&\approx E\left\{[\mathbf{w} - f(\boldsymbol{\mu}_\mathbf{u})][\mathbf{w} - f(\boldsymbol{\mu}_\mathbf{u})]^t\right\}, \\
&\approx E\left\{[J_f(\boldsymbol{\mu}_\mathbf{u})(\mathbf{u} - \boldsymbol{\mu}_\mathbf{u})][J_f(\boldsymbol{\mu}_\mathbf{u})(\mathbf{u} - \boldsymbol{\mu}_\mathbf{u})]^t\right\}, \\
&= J_f(\boldsymbol{\mu}_\mathbf{u}) E\left\{[(\mathbf{u} - \boldsymbol{\mu}_\mathbf{u})][(\mathbf{u} - \boldsymbol{\mu}_\mathbf{u})]^t\right\} J_f^t(\boldsymbol{\mu}_\mathbf{u}), \\
&= J_f(\boldsymbol{\mu}_\mathbf{u}) C_\mathbf{u} J_f^t(\boldsymbol{\mu}_\mathbf{u}).
\end{aligned} \tag{4-17}$$

Moreover, from the last expression, the inverse covariance matrix of $C_\mathbf{w}$ is given, to a first order approximation, by

$$C_\mathbf{w}^{-1} = J_f^{-t}(\boldsymbol{\mu}_\mathbf{u}) C_\mathbf{u}^{-1} J_f^{-1}(\boldsymbol{\mu}_\mathbf{u}), \tag{4-18}$$

provided $|J_f(\boldsymbol{\mu}_\mathbf{u})| \neq 0$.

## 5. FILTER RECURSIONS

The PKF is a linear filter that operates on debiased, converted measurements, $\bar{\mathbf{z}}_x(k)$, with precision matrices, $\bar{R}_x^{-1}(k)$, each derived from the complete measurement vector, $\mathbf{z}_z(k)$, with covariance matrices, $R_{\tilde{z}}(k)$, as described in sections 6 and 7, respectively. Once $\bar{\mathbf{z}}_x(k)$ and $\bar{R}_x^{-1}(k)$ are computed, the updates for the state and its covariance matrix can be computed using the standard Kalman information filter recursions (reference 1). Specifically, given the state estimate, $\hat{\mathbf{x}}_x(k-1|k-1)$, and its covariance matrix, $P_x(k-1|k-1)$, at time index, $k-1$, the PKF recursions for the updated state and its covariance matrix take the following form:

1. Compute the predicted state:
$$\hat{\mathbf{x}}_x(k|k-1) = A(k)\,\hat{\mathbf{x}}_x(k-1|k-1). \tag{5-1}$$

2. Compute the predicted state covariance matrix:
$$P_x(k|k-1) = A(k)\,P_x(k-1|k-1)\,A^t(k) + Q_x(k). \tag{5-2}$$

3. Compute the updated state covariance matrix:
$$P_x(k|k) = \left[P_x^{-1}(k|k-1) + \bar{R}_x^{-1}(k)\right]^{-1}. \tag{5-3}$$

4. Compute the Kalman gain matrix:
$$G(k) = P_x(k|k)\,\bar{R}_x^{-1}(k). \tag{5-4}$$

5. Compute the updated state estimate:
$$\hat{\mathbf{x}}_x(k|k) = \hat{\mathbf{x}}_x(k|k-1) + G(k)\left[\bar{\mathbf{z}}_x(k) - \hat{\mathbf{x}}_x(k|k-1)\right]. \tag{5-5}$$

In section 7, it is shown that the $N \times N$ converted measurement precision matrix, $\bar{R}_x^{-1}(k)$, has a $2 \times 2$, block-diagonal form comprised of $M \times M$ and $(N-M) \times (N-M)$ upper and lower diagonal blocks, respectively, and appropriately sized off-diagonal blocks. In certain cases, the three blocks comprising the last $N-M$ rows and columns of $\bar{R}_x^{-1}(k)$ contain all zeros; in this these cases, the last $N-M$ columns of the gain matrix, $G(k)$, will contain all zeros (by equation (5-4)) and, by equation (5-5), the last $N-M$ entries in the debiased, converted measurement vector, $\bar{\mathbf{z}}_x(k)$, will have no impact on the state update. An example of such a case is provided in section 10.

## 6. CONVERTED MEASUREMENT DEBIASING FUNCTION

The nonlinear filtering problem of interest in this report has measurements that are unbiased; indeed, from equation (3-2),
$$\mathrm{E}\left\{\mathbf{z}_{z_m}(k)\right\} = \mathrm{E}_{\tilde{\mathbf{z}}_{z_m}}\{h_m(\mathbf{x}_x(k)) + \tilde{\mathbf{z}}_{z_m}(k)\} = h_m(\mathbf{x}_x(k)), \tag{6-1}$$

where the last equality follows from the assumption that the measurement noise vector, $\tilde{\mathbf{z}}_{z_m}(k)$, has zero mean.

Likewise, it is desirable for the converted measurement, $\mathbf{z}_x(k) = g(\mathbf{z}_z(k))$, to be unbiased. As this is not likely to be the case in general, a debiasing function, $b : \mathcal{X} \to \mathcal{X}$, is introduced so that the converted measurement defined by

$$\bar{\mathbf{z}}_x(k) = b(\mathbf{z}_x(k)) = b(g[\mathbf{z}_z(k)]) = b(g[h(\mathbf{x}_x(k)) + \tilde{\mathbf{z}}_z(k)]), \tag{6-2}$$

is unbiased. The last equality in equation (6-2) follows from assumption 4. In order for $\bar{\mathbf{z}}_x(k)$ to be unbiased, the debiasing function must satisfy the following equality:

$$\mathrm{E}\left\{\bar{\mathbf{z}}_x(k)\right\} = \mathrm{E}_{\tilde{\mathbf{z}}_z}\{b(g[h(\mathbf{x}_x(k)) + \tilde{\mathbf{z}}_z(k)])\} = \mathrm{E}_{\tilde{\mathbf{z}}_z}\{g[h(\mathbf{x}_x(k))] + \tilde{\mathbf{z}}_z(k)\} = \mathbf{x}_x(k). \tag{6-3}$$

This equality defines an integral equation for the debiasing function, $b$, which may be written, in general, as

$$\mathbf{x}_x(k) = \int_{\mathbb{R}^N} b(g[h(\mathbf{x}_x(k)) + \tilde{\mathbf{z}}_z(k)]) \, p(\tilde{\mathbf{z}}_z(k)) \, d\tilde{\mathbf{z}}_z(k), \tag{6-4}$$

where $p$ is the probability density function (PDF) for the measurement noise vector, $\tilde{\mathbf{z}}_z(k)$. Under assumption 4, $p$ is the multivariate Gaussian PDF with mean zero and covariance matrix $\mathrm{R}_z$. Using the change of variables,

$$\mathbf{y}_x(k) = g[h(\mathbf{x}_x(k)) + \tilde{\mathbf{z}}_z(k)], \tag{6-5}$$

this integral equation may be rewritten as

$$\mathbf{x}_x(k) = \int_{\mathbb{R}^N} b(\mathbf{y}_x(k)) \, p(h(\mathbf{y}_x(k)) - h(\mathbf{x}_x(k))) \, |\mathbf{J}_h(\mathbf{y}_x(k))| \, d\mathbf{y}_x(k), \tag{6-6}$$

which, for $N = 1$, is a Fredholm (integral) equation of the first kind. Even in this limiting case, closed-form solutions for the unknown function, $b$, are available for only special cases. Nevertheless, Fredholm theory provides a potential starting point for the solution of equation (6-6) (see references 18 and 19). A general solution of equation (6-6) for the debiasing function, $b$, is beyond the scope of this report. Instead, this report considers two particular forms of the debiasing function, each with practical application.

## 6.1 ADDITIVE FORM

Suppose the debiasing function, $b$, is additive; that is,

$$b(\mathbf{z}_x(k)) = \mathbf{z}_x(k) + \mathbf{b}(k), \tag{6-7}$$

for some vector, $\mathbf{b}(k)$. Substituting equation (6-7) into equation (6-3) and solving for $\mathbf{b}(k)$ yields

$$\mathbf{b}(k) = \mathbf{x}_x(k) - \mathrm{E}_{\tilde{\mathbf{z}}_z}\{g[h(\mathbf{x}_x(k)) + \tilde{\mathbf{z}}_z(k)]\}. \tag{6-8}$$

In certain cases, the expectation in equation (6-8) can be evaluated analytically, and the unknown state, $\mathbf{x}_x(k)$, can be algebraically eliminated from the right-hand side, giving a closed-form expression for $\mathbf{b}(k)$. In the general case, an approximate solution for $\mathbf{b}(k)$ may be obtained by using the results of assumption 5 to replace terms in equation (6-8) involving the unknown state, $\mathbf{x}_x(k)$, with terms involving the predicted state, $\hat{\mathbf{x}}_x(k|k-1)$, and the error term, $\tilde{\mathbf{x}}_z(k|k-1)$, and taking an additional expectation of the resulting expression with respect to $\tilde{\mathbf{x}}_z(k|k-1)$. The resulting approximation is written explicitly as

$$\mathbf{b}(k) \approx \mathrm{E}_{\tilde{\mathbf{x}}_z}\{g[h(\hat{\mathbf{x}}_x(k|k-1)) - \tilde{\mathbf{x}}_z(k|k-1)]\}$$
$$- \mathrm{E}_{\tilde{\mathbf{x}}_z}\{\mathrm{E}_{\tilde{\mathbf{z}}_z}\{g[h(\hat{\mathbf{x}}_x(k|k-1)) - \tilde{\mathbf{x}}_z(k|k-1) + \tilde{\mathbf{z}}_z(k)]\}\}, \tag{6-9}$$
$$= \mathrm{E}_{\tilde{\mathbf{x}}_z}\{g[h(\hat{\mathbf{x}}_x(k|k-1)) - \tilde{\mathbf{x}}_z(k|k-1)]\} - \mathrm{E}_{\tilde{\mathbf{v}}_z}\{g[h(\hat{\mathbf{x}}_x(k|k-1)) - \tilde{\mathbf{v}}_z(k)]\}. \tag{6-10}$$

In general, numerical integration methods are required to compute the expectations in equation (6-10). Because the random vectors, $\tilde{\mathbf{x}}_z(k|k-1)$ and $\tilde{\mathbf{v}}_z(k)$, are multivariate Gaussian distributed with zero means and known covariance matrices, the Gaussian cubature methods developed in reference 20 are particularly attractive for this problem. For a more in depth discussion and additional numerical integration methods see Chapter 5 in reference 21.

## 6.2 MULTIPLICATIVE FORM

Suppose the debiasing function, $b$, is multiplicative; that is,

$$b(\mathbf{z}_x(k)) = \mathrm{B}(k)\,\mathbf{z}_x(k), \tag{6-11}$$

for some matrix, $\mathrm{B}(k)$. Substituting equation (6-11) into equation (6-3) yields

$$\mathrm{B}(k)\,\mathrm{E}_{\tilde{\mathbf{z}}_z}\{g[h(\mathbf{x}_x(k)) + \tilde{\mathbf{z}}_z(k)]\} = \mathbf{x}_x(k). \tag{6-12}$$

As for the additive form for $b(k)$, in certain cases, the expectation in equation (6-12) can be evaluated symbolically, and the unknown state, $\mathbf{x}_x(k)$, can be algebraically eliminated from the right-hand side, giving a closed-form expression for $\mathrm{B}(k)$.

Of particular interest in some practical applications is the case where the debiasing matrix, $\mathrm{B}(k)$, is diagonal. Let $\mathrm{diag}(\mathbf{x}_x(k))$ denote the diagonal matrix formed from the vector, $\mathbf{x}_x(k)$, and let "$\circ$" denote the Hadamard product (or element-wise product) of two matrices. Then,

$$\mathrm{B}(k) = \mathrm{diag}(\mathrm{E}_{\tilde{\mathbf{z}}_z}\{g[h(\mathbf{x}_x(k)) + \tilde{\mathbf{z}}_z(k)]\})^{-1} \circ \mathrm{diag}(\mathbf{x}_x(k)), \tag{6-13}$$

provided the diagonal matrix involving the expectation in this equation is invertible. In the case where the unknown state, $\mathbf{x}_x(k)$, cannot be eliminated from equation (6-13), assumption 5 may be used, as in section 6.1, to approximate the solution for the debiasing matrix, $\mathrm{B}(k)$, by replacing terms in equation (6-13), involving $\mathbf{x}_x(k)$, with terms involving $\hat{\mathbf{x}}_x(k|k-1)$ and $\tilde{\mathbf{x}}_z(k|k-1)$, and taking an additional expectation of the resulting vector with respect to $\tilde{\mathbf{x}}_z(k|k-1)$. The resulting approximation is written explicitly as

$$\mathrm{B}(k) \approx \mathrm{diag}\left(\mathrm{E}_{\tilde{\mathbf{x}}_z}\{\mathrm{E}_{\tilde{\mathbf{z}}_z}\{g[h(\hat{\mathbf{x}}_x(k|k-1)) - \tilde{\mathbf{x}}_z(k|k-1) + \tilde{\mathbf{z}}_z(k)]\}\}\right)^{-1}$$
$$\circ \mathrm{diag}\left(\mathrm{E}_{\tilde{\mathbf{x}}_z}\{g[h(\hat{\mathbf{x}}_x(k|k-1)) - \tilde{\mathbf{x}}_z(k|k-1)]\}\right) \tag{6-14}$$
$$= \mathrm{diag}\left(\mathrm{E}_{\tilde{\mathbf{v}}_z}\{g[h(\hat{\mathbf{x}}_x(k|k-1)) - \tilde{\mathbf{v}}_z(k)\}]\right)^{-1}$$
$$\circ \mathrm{diag}\left(\mathrm{E}_{\tilde{\mathbf{x}}_z}\{g[h(\hat{\mathbf{x}}_x(k|k-1)) - \tilde{\mathbf{x}}_z(k|k-1)]\}\right). \tag{6-15}$$

In general, numerical integration methods such as those described in references 20 and 21 are required to compute the expectations in equation (6-15). The numerical methods used to evaluate the expectations in the PKF are described in section 8.

# 7. CONVERTED MEASUREMENT PRECISION MATRIX

By definition, the covariance matrix of the debiased, converted measurement, $\bar{\mathbf{z}}_x(k)$, defined by equation (6-2), is given by

$$\begin{aligned}\mathbf{R}_x(k) &= \mathrm{E}\left\{[\bar{\mathbf{z}}_x(k) - \mathrm{E}\{\bar{\mathbf{z}}_x(k)\}][\bar{\mathbf{z}}_x(k) - \mathrm{E}\{\bar{\mathbf{z}}_x(k)\}]^t\right\}, \\ &= \mathrm{E}\left\{[\bar{\mathbf{z}}_x(k) - \mathbf{x}_x(k)][\bar{\mathbf{z}}_x(k) - \mathbf{x}_x(k)]^t\right\},\end{aligned} \quad (7\text{-}1)$$

where the the second line of this equation follows from equation (6-3). Expanding the quadratic in equation (7-1), simplifying the resulting expression, and writing $\bar{\mathbf{z}}_x(k)$ in terms of $\mathbf{x}_x(k)$, as in equation (6-2), yields

$$\mathbf{R}_x(k) = \mathrm{E}_{\tilde{\mathbf{z}}_z}\left\{b(g\left[h(\mathbf{x}_x(k)) + \tilde{\mathbf{z}}_z(k)\right])\,b^t(g\left[h(\mathbf{x}_x(k)) + \tilde{\mathbf{z}}_z(k)\right])\right\} - \mathbf{x}_x(k)\,\mathbf{x}_x^t(k), \quad (7\text{-}2)$$

where it is made explicit that the expectation is with respect to the measurement noise, $\tilde{\mathbf{z}}_z(k)$.

Even if the expectation in equation (7-2) can be evaluated analytically, there are two problems with this expression for $\mathbf{R}_x(k)$ that make it impractical. First, in general, the right-hand side of equation (7-2) is a function of the unknown state, $\mathbf{x}_x(k)$. While some authors have sought to circumvent this problem in specific applications by replacing $\mathbf{x}_x(k)$ in equation (7-2) with an appropriate transformation of the measurement, $\mathbf{z}_z(k)$ (reference 13), or the predicted state, $\hat{\mathbf{x}}_x(k|k-1)$ (references 8, 22), both approaches lead to either a biased or inconsistent state estimate. Second, in the case $M < N$, equation (7-2) does not compensate for the information added by the unobserved measurement, $\mathbf{z}_{z_u}(k)$, which, together with the observed measurement, $\mathbf{z}_{z_m}(k)$, are required to compute the debiased, converted measurement, $\bar{\mathbf{z}}_x(k)$. The remainder of this section is devoted to the derivation of the converted measurement precision matrix, $\bar{\mathbf{R}}_x^{-1}(k)$, which addresses both problems, and is used, in conjunction with the debiased, converted measurement, $\bar{\mathbf{z}}_x(k)$, in the information filter recursions for the state estimate, $\hat{\mathbf{x}}_x(k|k)$, listed in section 5.

The first step in the derivation of $\bar{\mathbf{R}}_x^{-1}(k)$ is to rewrite equation (7-2) for $\mathbf{R}_x(k)$ in terms of the state in measurement coordinates, i.e., $\mathbf{x}_z(k)$. The following equalities follow from the definitions of the mappings $h$ and $g$, and assumption 1:

$$h(\mathbf{x}_x(k)) = \mathbf{x}_z(k), \quad (7\text{-}3)$$

$$\mathbf{x}_x(k) = g[h(\mathbf{x}_x(k))] = g[\mathbf{x}_z(k)]. \quad (7\text{-}4)$$

Substituting the right-hand sides of these equalities into equation (7-2) yields

$$\mathbf{R}_x(k) = \mathrm{E}_{\tilde{\mathbf{z}}_z}\left\{b(g\left[\mathbf{x}_z(k) + \tilde{\mathbf{z}}_z(k)\right])\,b^t(g\left[\mathbf{x}_z(k) + \tilde{\mathbf{z}}_z(k)\right])\right\} - g[\mathbf{x}_z(k)]\,g^t[\mathbf{x}_z(k)]. \quad (7\text{-}5)$$

In this form, the converted measurement covariance matrix, $\mathbf{R}_x(k)$, may be estimated using the predicted state, $\hat{\mathbf{x}}_x(k|k-1)$, and assumption 5. Specifically, substituting the right-hand side of the approximation for $\mathbf{x}_z(k)$, given by equation (4-4), into equation (7-5), and taking an additional expectation over the predicted state error, $\tilde{\mathbf{x}}_z(k|k-1)$, gives the estimate

$$\begin{aligned}\hat{\mathbf{R}}_x(k) = {}& \mathrm{E}_{\tilde{\mathbf{x}}_z}\Big\{\mathrm{E}_{\tilde{\mathbf{z}}_z}\big\{b(g\left[h(\hat{\mathbf{x}}_x(k|k-1)) - \tilde{\mathbf{x}}_z(k|k-1) + \tilde{\mathbf{z}}_z(k)\right]) \\ & \times b^t(g\left[h(\hat{\mathbf{x}}_x(k|k-1)) - \tilde{\mathbf{x}}_z(k|k-1) + \tilde{\mathbf{z}}_z(k)\right])\big\}\Big\} \\ & - \mathrm{E}_{\tilde{\mathbf{x}}_z}\big\{g[h(\hat{\mathbf{x}}_x(k|k-1)) - \tilde{\mathbf{x}}_z(k|k-1)]\,g^t[h(\hat{\mathbf{x}}_x(k|k-1)) - \tilde{\mathbf{x}}_z(k|k-1)]\big\}.\end{aligned} \quad (7\text{-}6)$$

Furthermore, using the change of variables defined by equation (4-15), the double expectation in equation (7-6) can be reduced to a single expectation over the combined error, $\tilde{\mathbf{v}}_z(k)$, yielding

$$\hat{\mathrm{R}}_x(k) = \mathrm{E}_{\tilde{\mathbf{v}}_z}\left\{b(g\left[h(\hat{\mathbf{x}}_x(k|k-1)) - \tilde{\mathbf{v}}_z(k)\right])\, b^t(g\left[h(\hat{\mathbf{x}}_x(k|k-1)) - \tilde{\mathbf{v}}_z(k)\right])\right\} \\ - \mathrm{E}_{\tilde{\mathbf{x}}_z}\left\{g[h(\hat{\mathbf{x}}_x(k|k-1)) - \tilde{\mathbf{x}}_z(k|k-1)]\, g^t[h(\hat{\mathbf{x}}_x(k|k-1)) - \tilde{\mathbf{x}}_z(k|k-1)]\right\}. \quad (7\text{-}7)$$

In the case where the debiasing function, $b$, is additive, as in equation (6-7), the estimated converted measurement covariance matrix becomes

$$\hat{\mathrm{R}}_x(k) = \mathrm{E}_{\tilde{\mathbf{v}}_z}\left\{g\left[h(\hat{\mathbf{x}}_x(k|k-1)) - \tilde{\mathbf{v}}_z(k)\right]\, g^t\left[h(\hat{\mathbf{x}}_x(k|k-1)) - \tilde{\mathbf{v}}_z(k)\right]\right\} + \mathbf{b}(k)\mathbf{b}^t(k) \\ + \mathrm{E}_{\tilde{\mathbf{v}}_z}\left\{g\left[h(\hat{\mathbf{x}}_x(k|k-1)) - \tilde{\mathbf{v}}_z(k)\right]\right\}\mathbf{b}^t(k) \\ + \mathbf{b}(k)\,\mathrm{E}_{\tilde{\mathbf{v}}_z}\left\{g^t\left[h(\hat{\mathbf{x}}_x(k|k-1)) - \tilde{\mathbf{v}}_z(k)\right]\right\} \\ - \mathrm{E}_{\tilde{\mathbf{x}}_z}\left\{g[h(\hat{\mathbf{x}}_x(k|k-1)) - \tilde{\mathbf{x}}_z(k|k-1)]\, g^t[h(\hat{\mathbf{x}}_x(k|k-1)) - \tilde{\mathbf{x}}_z(k|k-1)]\right\}. \quad (7\text{-}8)$$

Likewise, in the case where $b$ is multiplicative, as in equation (6-11), equation (7-7) becomes

$$\hat{\mathrm{R}}_x(k) = \mathrm{B}(k)\, \mathrm{E}_{\tilde{\mathbf{v}}_z}\left\{g\left[h(\hat{\mathbf{x}}_x(k|k-1)) - \tilde{\mathbf{v}}_z(k)\right]\, g^t\left[h(\hat{\mathbf{x}}_x(k|k-1)) - \tilde{\mathbf{v}}_z(k)\right]\right\}\mathrm{B}(k) \\ - \mathrm{E}_{\tilde{\mathbf{x}}_z}\left\{g[h(\hat{\mathbf{x}}_x(k|k-1)) - \tilde{\mathbf{x}}_z(k|k-1)]\, g^t[h(\hat{\mathbf{x}}_x(k|k-1)) - \tilde{\mathbf{x}}_z(k|k-1)]\right\}. \quad (7\text{-}9)$$

As for the expectations required to compute the debiasing function discussed section 6, numerical methods are, in general, required to compute the expectations in equations (7-8) and (7-9). Again, since $\tilde{\mathbf{x}}_z(k|k-1)$ and $\tilde{\mathbf{v}}_z(k)$ are multivariate Gaussian distributed with zero means and known covariance matrices, Gaussian cubature methods are natural for this problem.

The next step in the derivation of $\bar{\mathrm{R}}_x^{-1}(k)$ is to remove the information added by the unobserved measurement, $\mathbf{z}_{z_u}(k)$, from the estimated converted measurement precision matrix, $\hat{\mathrm{R}}_x^{-1}(k)$. When the state and measurement coordinate systems have the same dimension, that is, when $M = N$, there is no need to introduce the unobserved measurement, $\mathbf{z}_{z_u}(k)$, and

$$\bar{\mathrm{R}}_x^{-1}(k) = \hat{\mathrm{R}}_x^{-1}(k). \quad (7\text{-}10)$$

However, when $M < N$, the information content of the precision matrix, $\hat{\mathrm{R}}_x^{-1}(k)$, is influenced by the unobserved measurement covariance matrix, $\mathrm{R}_{z_u}(k)$, and the observed/unobserved measurement cross-covariance matrix, $\mathrm{R}_{z_m z_u}(k)$, respectively, because the covariance matrix of the error term, $\tilde{\mathbf{v}}_z(k)$, in equations (7-8) and (7-9) is equal to $\mathrm{P}_z(k|k-1) + \mathrm{R}_z(k)$ (see assumption 6). To mitigate this influence, the precision matrix, $\hat{\mathrm{R}}_x^{-1}(k)$, is first mapped to measurement coordinates, where its last $N - M$ rows and columns are set to zeros, and then mapped back to state coordinates, effectively zeroing the information contribution of the unobserved measurement, $\mathbf{z}_{z_u}(k)$, to the converted measurement precision matrix, $\hat{\mathrm{R}}_x^{-1}(k)$. In equation form, this transformation and information zeroing procedure is expressed compactly as

$$\bar{\mathrm{R}}_x^{-1}(k) = \mathrm{J}_g^{-t}(k)\left[\mathrm{W}\left(\mathrm{J}_h^{-t}(k)\,\hat{\mathrm{R}}_x^{-1}(k)\,\mathrm{J}_h^{-1}(k)\right)\mathrm{W}^t\right]\mathrm{J}_g^{-1}(k), \quad (7\text{-}11)$$

where W is a $2 \times 2$ block matrix with total dimension $N \times N$, with the upper left block equal to the $M \times M$ identity matrix, denoted $\mathrm{I}_{M \times M}$, and all other blocks equal to appropriately dimensioned matrices of all zeros; that is,

$$
\mathbf{W} = \begin{bmatrix} & & & 0 & \cdots & 0 \\ \mathbf{I}_{M \times M} & & & \vdots & \ddots & \vdots \\ & & & 0 & \cdots & 0 \\ 0 & \cdots & 0 & 0 & \cdots & 0 \\ \vdots & \ddots & \vdots & \vdots & \ddots & \vdots \\ 0 & \cdots & 0 & 0 & \cdots & 0 \end{bmatrix}, \tag{7-12}
$$

and $\mathbf{J}_g(k)$ and $\mathbf{J}_h(k)$ are shorthand notations for the Jacobian matrices, $\mathbf{J}_g(\hat{\mathbf{x}}_z(k|k-1))$ and $\mathbf{J}_h(\hat{\mathbf{x}}_x(k|k-1))$, respectively. When $M = N$, the matrix, $\mathbf{W}$, equals the $N \times N$ identify matrix and, by virtue of assumption 2, equation (7-11) reduces to equation (7-10), as expected.

Finally, when $M < N$, it is emphasized that the information zeroing matrix, $W$, has rank, $M$, and, hence, the converted measurement precision matrix, $\bar{\mathbf{R}}_x^{-1}(k)$, is not full rank. In this case, it is instructive, for certain analyses, to partition the matrix, $\mathbf{J}_g^{-t}(k)$, as

$$
\mathbf{J}_g^{-t}(k) = \begin{bmatrix} \mathbf{J}_{g,m}^{-t}(k) & \mathbf{J}_{g,mu}^{-t}(k) \\ \mathbf{J}_{g,um}^{-t}(k) & \mathbf{J}_{g,u}^{-t}(k) \end{bmatrix}, \tag{7-13}
$$

where $\mathbf{J}_{g,m}^{-t}(k)$ and $\mathbf{J}_{g,u}^{-t}(k)$ correspond to the first $M \times M$ and last $(N-M) \times (N-M)$ diagonal blocks of $\mathbf{J}_g^{-t}(k)$, respectively, and $\mathbf{J}_{g,mu}^{-t}(k)$ and $\mathbf{J}_{g,um}^{-t}(k)$ correspond to the appropriate off-diagonal blocks of $\mathbf{J}_g^{-t}(k)$. By defining the converted measurement precision matrix in measurement coordinates, $\bar{\mathbf{R}}_z^{-1}(k)$, as the term in square brackets in equation (7-11), and partitioning $\bar{\mathbf{R}}_z^{-1}(k)$ as

$$
\bar{\mathbf{R}}_z^{-1}(k) = \begin{bmatrix} & & & 0 & \cdots & 0 \\ \bar{\mathbf{R}}_{z_m}^{-1}(k) & & & \vdots & \ddots & \vdots \\ & & & 0 & \cdots & 0 \\ 0 & \cdots & 0 & 0 & \cdots & 0 \\ \vdots & \ddots & \vdots & \vdots & \ddots & \vdots \\ 0 & \cdots & 0 & 0 & \cdots & 0 \end{bmatrix}, \tag{7-14}
$$

where $\bar{\mathbf{R}}_{z_m}^{-1}(k)$ denotes the upper left, $M \times M$ block of $\bar{\mathbf{R}}_z^{-1}(k)$, it follows, using equation (7-13) and equation (7-14), the converted measurement precision matrix can be written in block form:

$$
\bar{\mathbf{R}}_x^{-1}(k) = \begin{bmatrix} \mathbf{J}_{g,m}^{-t}(k)\bar{\mathbf{R}}_{z_m}^{-1}(k)\mathbf{J}_{g,m}^{-1}(k) & \mathbf{J}_{g,m}^{-t}(k)\bar{\mathbf{R}}_{z_m}^{-1}(k)\mathbf{J}_{g,um}^{-1}(k) \\ \mathbf{J}_{g,um}^{-t}(k)\bar{\mathbf{R}}_{z_m}^{-1}(k)\mathbf{J}_{g,m}^{-1}(k) & \mathbf{J}_{g,um}^{-t}(k)\bar{\mathbf{R}}_{z_m}^{-1}(k)\mathbf{J}_{g,um}^{-1}(k) \end{bmatrix}. \tag{7-15}
$$

This block form of $\bar{\mathbf{R}}_x^{-1}(k)$ will be useful to the analysis of the examples presented in section 10.

# 8. NUMERICAL METHODS FOR EVALUATING EXPECTATIONS

In general, the expectations required by the converted measurement debiasing function and precision matrix cannot be evaluated analytically, and must be computed using suitable numerical integration methods. Based on its common usage in the target tracking literature, the term sigma point transform (SPT) will be used as a generic term for the unscented transform and Gaussian cubature which form the bases of the UKF in references 2, 4–6, 12, and cubature Kalman filter (CKF) in references 2, 20, 23, respectively. As stated previously, a more complete discussion of multidimensional numerical integration methods is given in Chapter 5 of reference 21. The accuracy of any numerical integration method is an important issue in filtering problems and should not be ignored. The discussion in section IV of reference 20 and chapter 5, section 10 of reference 21 provide some recommendations and insight.

To simplify notation for the analysis of this section, let

$$y(\mathbf{u}(k)) = g[h(\hat{\mathbf{x}}_x(k|k-1)) - \mathbf{u}(k)], \tag{8-1}$$

where $\mathbf{u}(k)$ may equal either $\tilde{\mathbf{x}}_z(k|k-1)$ or $\tilde{\mathbf{v}}_z(k)$, both of which are, by assumption 6, approximately Gaussian distributed with zero means and covariance matrices, $\mathbf{P}_z(k|k-1)$ and $\mathbf{P}_z(k|k-1) + \mathbf{R}_z(k)$, respectively.

Assuming either the additive or multiplicative form for the debiasing function, $b$, given by equations (6-7) and (6-11), respectively, the following expectations are required to compute the debiasing function parameter—$\mathbf{b}(k)$ in the case of the additive form, or $\mathbf{B}(k)$ in the case of the multiplicative form—and the precision matrix, $\bar{\mathbf{R}}_x^{-1}(k)$:

$$E_{\mathbf{u}}\{y(\mathbf{u}(k))\}, \tag{8-2}$$

$$E_{\mathbf{u}}\{y(\mathbf{u}(k))y^t(\mathbf{u}(k))\}; \tag{8-3}$$

see equations (6-10) and (6-15) and equations (7-8) and (7-9). The SPT approximates the mean, $\boldsymbol{\mu}_{\mathbf{u}}(k)$, and covariance, $\mathbf{C}_{\mathbf{u}}(k)$, of the nonlinear transformation, $y(\mathbf{u}(k))$, defined by

$$\boldsymbol{\mu}_{\mathbf{u}}(k) = E_{\mathbf{u}}\{y(\mathbf{u}(k))\}, \tag{8-4}$$

$$\mathbf{C}_{\mathbf{u}}(k) = E_{\mathbf{u}}\{y(\mathbf{u}(k))y^t(\mathbf{u}(k))\} - \boldsymbol{\mu}_{\mathbf{u}}(k)\boldsymbol{\mu}_{\mathbf{u}}^t(k), \tag{8-5}$$

respectively, with sample statistics derived from a finite set, $\mathscr{S}_{\mathbf{u}}(k) = \{\mathbf{u}^i(k)\}$, of $S \geq 1$ sigma points, $\mathbf{u}^i(k)$, with corresponding, real-valued weights, $\mathscr{W} = \{w_i\}$, as follows:

1. Pass the sigma points, $\mathscr{S}_{\mathbf{u}}(k)$, through the nonlinear transformation, $y$, to obtain the transformed set of sigma points, $\mathscr{Y}_{\mathbf{u}}(k) = \{\mathbf{y}_{\mathbf{u}}^i(k)\}$, with

$$\mathbf{y}_{\mathbf{u}}^i(k) = y(\mathbf{u}^i(k)). \tag{8-6}$$

2. Compute the weighted sample mean of the transformed sigma points,

$$\bar{\boldsymbol{\mu}}_{\mathbf{u}}(k) = \sum_{i=1}^{S} w_i\, \mathbf{y}_{\mathbf{u}}^i(k). \tag{8-7}$$

3. Compute the weighted sample covariance of the transformed sigma points,

$$\bar{C}_{\mathbf{u}}(k) = \sum_{i=1}^{S} w_i \left[\mathbf{y}_{\mathbf{u}}^i(k) - \bar{\boldsymbol{\mu}}_{\mathbf{u}}(k)\right] \left[\mathbf{y}_{\mathbf{u}}^i(k) - \bar{\boldsymbol{\mu}}_{\mathbf{u}}(k)\right]^t. \tag{8-8}$$

Substituting the sample statistics, $\bar{\boldsymbol{\mu}}_{\mathbf{u}}(k)$ and $\bar{C}_{\mathbf{u}}(k)$, for $\boldsymbol{\mu}_{\mathbf{u}}(k)$ and $C_{\mathbf{u}}(k)$, respectively, in equations (8-4) and (8-5), and solving for the desired expectations in equations (8-2) and (8-3), yields the approximations

$$E_{\mathbf{u}}\{y(\mathbf{u}(k))\} \approx \bar{\boldsymbol{\mu}}_{\mathbf{u}}(k), \tag{8-9}$$

$$E_{\mathbf{u}}\{y(\mathbf{u}(k))y^t(\mathbf{u}(k))\} \approx \bar{C}_{\mathbf{u}}(k) + \bar{\boldsymbol{\mu}}_{\mathbf{u}}(k)\bar{\boldsymbol{\mu}}_{\mathbf{u}}^t(k). \tag{8-10}$$

Choice of the sigma points, $\mathscr{S}_{\mathbf{u}}(k)$, and weights, $\mathscr{W}$, depends on the particular transform (e.g., unscented, cubature, etc.). See references 2, 4–6, 12 and 2, 20, 21, 23 for details on how to compute these sets for the unscented and cubature transforms, respectively.

To insure the approximations in equations (8-9) and (8-10) are appropriate, one should evaluate these expectations offline using a sequence of increasingly accurate SPTs and choose the smallest SPT that meets accuracy requirements. Sometimes it is necessary to implement the filter using different order SPTs and choose the SPT that gives the best performance. In practice, performance is a trade off between computational complexity and numerical accuracy.

## 9. CONVERTED MEASUREMENT COMPUTATIONS

The filter recursions listed in section 5 require computations of the converted measurement, $\bar{\mathbf{z}}_x(k)$, and converted measurement precision matrix, $\bar{R}_x^{-1}(k)$, the latter of which, given by equation (7-11), is a transformation of the estimated precision matrix, $\hat{R}_x^{-1}(k)$.

The computations of $\bar{\mathbf{z}}_x(k)$ and $\hat{R}_x^{-1}(k)$ require specification of the form of the debiasing function, $b$, and evaluations of the expectations given by equations (8-2) and (8-3) for both $\mathbf{u}(k) = \tilde{\mathbf{x}}_z(k|k-1)$ and $\mathbf{u}(k) = \tilde{\mathbf{v}}_z(k)$. These expectations, in turn, may be approximated using equations (8-9) and (8-10) and the sample means, $\bar{\boldsymbol{\mu}}_{\tilde{\mathbf{x}}_z}(k)$ and $\bar{\boldsymbol{\mu}}_{\tilde{\mathbf{v}}_z}(k)$, and sample covariances, $\bar{C}_{\tilde{\mathbf{x}}_z}(k)$ and $\bar{C}_{\tilde{\mathbf{v}}_z}(k)$, computed from the sigma point sets,

$$\mathscr{S}_{\tilde{\mathbf{x}}_z}(k) = \left\{\tilde{\mathbf{x}}_z^i(k|k-1) : i = 1, \ldots, S\right\} \tag{9-1}$$

and

$$\mathscr{S}_{\tilde{\mathbf{v}}_z}(k) = \left\{\tilde{\mathbf{v}}_z^i(k) : i = 1, \ldots, S\right\}, \tag{9-2}$$

using the procedure discussed in the section 8. Sections 9.1 and 9.2 derive explicit expressions for $\bar{\mathbf{z}}_x(k)$ and $\hat{R}_x^{-1}(k)$ for the additive and multiplicative forms of the debiasing function, $b$, given in sections 6.1 and 6.2, respectively, in terms of the SPT sample means, $\bar{\boldsymbol{\mu}}_{\tilde{\mathbf{x}}_z}(k)$ and $\bar{\boldsymbol{\mu}}_{\tilde{\mathbf{v}}_z}(k)$, and sample covariances, $\bar{C}_{\tilde{\mathbf{x}}_z}(k)$ and $\bar{C}_{\tilde{\mathbf{v}}_z}(k)$.

## 9.1 ADDITIVE DEBIASING FUNCTION

From equations (6-2) and (6-7), the converted measurement for the additive form of the debiasing function is given by

$$\bar{\mathbf{z}}_x(k) = g[\mathbf{z}_z(k)] + \mathbf{b}(k), \tag{9-3}$$

where debiasing vector, $\mathbf{b}(k)$, is approximated by equation (6-10). Using equation (8-9) to evaluate the expectations in equation (6-10) yields

$$\mathbf{b}(k) = \bar{\boldsymbol{\mu}}_{\tilde{\mathbf{x}}_z}(k) - \bar{\boldsymbol{\mu}}_{\tilde{\mathbf{v}}_z}(k). \tag{9-4}$$

Substituting this result into equation (7-8), using equations (8-9) and (8-10) to evaluate the remaining expectations in equation (7-8), and simplifying the result yields

$$\hat{\mathbf{R}}_x(k) = \bar{\mathbf{C}}_{\tilde{\mathbf{v}}_z}(k) - \bar{\mathbf{C}}_{\tilde{\mathbf{x}}_z}(k). \tag{9-5}$$

## 9.2 MULTIPLICATIVE DEBIASING FUNCTION

From equations (6-2) and (6-11), the converted measurement for the multiplicative form of the debiasing function is given by

$$\bar{\mathbf{z}}_x(k) = \mathbf{B}(k) g[\mathbf{z}_z(k)], \tag{9-6}$$

where the debiasing matrix, $\mathbf{B}(k)$, is approximated by equation (6-15). Using equation (8-9) to evaluate the expectations in equation (6-15) yields

$$\mathbf{B}(k) = \text{diag}\left(\bar{\boldsymbol{\mu}}_{\tilde{\mathbf{v}}_z}(k)\right)^{-1} \circ \text{diag}\left(\bar{\boldsymbol{\mu}}_{\tilde{\mathbf{x}}_z}(k)\right). \tag{9-7}$$

Likewise, using equation (8-10) to evaluate the expectations in equation (7-9) yields

$$\hat{\mathbf{R}}_x(k) = \mathbf{B}(k)\left[\bar{\mathbf{C}}_{\tilde{\mathbf{v}}_z}(k) + \bar{\boldsymbol{\mu}}_{\tilde{\mathbf{v}}_z}(k)\bar{\boldsymbol{\mu}}^t_{\tilde{\mathbf{v}}_z}(k)\right]\mathbf{B}(k) - \left[\bar{\mathbf{C}}_{\tilde{\mathbf{x}}_z}(k) + \bar{\boldsymbol{\mu}}_{\tilde{\mathbf{x}}_z}(k)\bar{\boldsymbol{\mu}}^t_{\tilde{\mathbf{x}}_z}(k)\right]. \tag{9-8}$$

Finally, substituting equation (9-7) into equation (9-8) and simplifying the result yields

$$\hat{\mathbf{R}}_x(k) = \mathbf{B}(k)\bar{\mathbf{C}}_{\tilde{\mathbf{v}}_z}(k)\mathbf{B}(k) - \bar{\mathbf{C}}_{\tilde{\mathbf{x}}_z}(k). \tag{9-9}$$

## 10. EXAMPLE: POLAR-TO-CARTESIAN COORDINATE CONVERSION

A practical example from the target tracking literature is presented in this section; namely, the conversion of polar measurements to Cartesian coordinates, where the polar measurements may include any subset of target range, bearing, and range rate measurements. In this example, the converted measurement debiasing function can be derived analytically, but the expectations required by the converted measurement precision matrix must be evaluated numerically. In general, neither the debiasing function nor the precision matrix can be derived analytically, and numerical methods such as those discussed in section 8 must be used to evaluate all the required expectations. The bistatic tracking problem described in reference 10 converts bistatic range and bearing measurements to Cartesian coordinates and is an example of the general case. Application of the filtering method developed in this report to the bistatic tracking problem is left for future work.

In the polar-to-Cartesian coordinate conversion example, measurements of target range, bearing, and/or range rate are assumed, the last of which is equivalent to relative velocity along the line of sight. For use in the PKF, these measurements are augmented with a non-informative, unobserved measurement of cross-range rate, or relative velocity across the line of sight, as shown in the following and discussed in reference 9. For brevity, the time index, $k$, is dropped in this section.

## 10.1 STATE AND MEASUREMENT COORDINATE MAPPINGS

For simplicity, assume a stationary sensor at the origin of a planar Cartesian coordinate system. Let $r$ and $\alpha$ denote the true range and bearing to the target, respectively, where $\alpha$ is measured counterclockwise from the positive $x$-axis. The true states of an assumed constant-velocity target in state and measurement coordinates for this problem are given by

$$\mathbf{x}_\mathcal{X} = \begin{bmatrix} x \\ y \\ \dot{x} \\ \dot{y} \end{bmatrix}, \quad \mathbf{x}_\mathcal{Z} = \begin{bmatrix} r \\ \alpha \\ \dot{r} \\ \dot{c} \end{bmatrix}, \tag{10-1}$$

respectively, where $\dot{c} = r\dot{\alpha}$ denotes the target cross-range rate; that is, the target's velocity across the line of sight in measurement coordinates. In this problem, the target's cross-range rate, $\dot{c}$, is not observed. One or more of the other measurement coordinates may be observed. Common examples of observed measurements are bearings-only, range and bearing, and range, bearing, and range rate. Consequently, the observed measurement coordinates, $\mathcal{Z}_m \subseteq \mathbb{R}^M$, $M \in \{1, 2, 3\}$, comprise the set of all possible target ranges, bearings, and/or range rates, while the unobserved measurement coordinates, $\mathcal{Z}_u \subseteq \mathbb{R}^{4-M}$, comprise the set of all possible target polar coordinates, in addition to cross-range rate, that are not observed.

The mapping, $g : \mathcal{Z} \to \mathcal{X}$, from the complete measurement space, $\mathcal{Z} = \mathcal{Z}_m \times \mathcal{Z}_u$, to the target state space, $\mathcal{X}$, is given by

$$\mathbf{x}_\mathcal{X} = g(\mathbf{x}_\mathcal{Z}) = \begin{bmatrix} r\cos\alpha \\ r\sin\alpha \\ \dot{r}\cos\alpha - \dot{c}\sin\alpha \\ \dot{r}\sin\alpha + \dot{c}\cos\alpha \end{bmatrix}, \tag{10-2}$$

where the last two entries are derived from the first two entries by differentiating the latter with respect to time. Likewise, the mapping, $h : \mathcal{X} \to \mathcal{Z}$, from the state space to the measurement space is given by

$$\mathbf{x}_\mathcal{Z} = h(\mathbf{x}_\mathcal{X}) = \begin{bmatrix} (x^2 + y^2)^{1/2} \\ \tan^{-1}(y/x) \\ (x\dot{x} + y\dot{y})/(x^2 + y^2)^{1/2} \\ (x\dot{y} - y\dot{x})/(x^2 + y^2)^{1/2} \end{bmatrix}, \tag{10-3}$$

where, again, the last two entries are derived from the first two entries by differentiating the latter with respect to time.

## 10.2 CONVERTED MEASUREMENT DEBIASING FUNCTION

Based on the analysis in references 8, 9, 13, the debiasing function, $b : \mathcal{X} \to \mathcal{X}$, as defined by equation (6-3), is multiplicative, so that $b(g(\mathbf{z}_z)) = Bg(\mathbf{z}_z)$ for some matrix B. Under this assumption, the matrix B is obtained as the solution to equation (6-12). Solving this equation for B requires evaluating the expected value of the raw converted measurement, $g(\mathbf{z}_z) = g(h(\mathbf{x}_x) + \tilde{\mathbf{z}}_z) = g(\mathbf{x}_z + \tilde{\mathbf{z}}_z)$, with respect to the measurement noise,

$$\tilde{\mathbf{z}}_z = \begin{bmatrix} w_r \\ w_\alpha \\ w_{\dot{r}} \\ w_{\dot{c}} \end{bmatrix}, \tag{10-4}$$

which is assumed to be zero-mean, Gaussian distributed with covariance matrix

$$R_z = \begin{bmatrix} \sigma_r^2 & 0 & \rho\sigma_r\sigma_{\dot{r}} & 0 \\ 0 & \sigma_\alpha^2 & 0 & 0 \\ \rho\sigma_r\sigma_{\dot{r}} & 0 & \sigma_{\dot{r}}^2 & 0 \\ 0 & 0 & 0 & \sigma_{\dot{c}}^2 \end{bmatrix} \tag{10-5}$$

(see reference 9 or chapter 4 of reference 7). Using equations (10-1), (10-2), and (10-4), the measurement, $\mathbf{z}_z$, is written as

$$\mathbf{z}_z = \mathbf{x}_z + \tilde{\mathbf{z}}_z = \begin{bmatrix} r \\ \alpha \\ \dot{r} \\ \dot{c} \end{bmatrix} + \begin{bmatrix} w_r \\ w_\alpha \\ w_{\dot{r}} \\ w_{\dot{c}} \end{bmatrix} = \begin{bmatrix} z_r \\ z_\alpha \\ z_{\dot{r}} \\ z_{\dot{c}} \end{bmatrix}, \tag{10-6}$$

and the raw (biased) converted measurement is written as

$$g(\mathbf{z}_z) = \begin{bmatrix} (r + w_r)\cos(\alpha + w_\alpha) \\ (r + w_r)\sin(\alpha + w_\alpha) \\ (\dot{r} + w_{\dot{r}})\cos(\alpha + w_\alpha) - (\dot{c} + w_{\dot{c}})\sin(\alpha + w_\alpha) \\ (\dot{r} + w_{\dot{r}})\sin(\alpha + w_\alpha) + (\dot{c} + w_{\dot{c}})\cos(\alpha + w_\alpha) \end{bmatrix}. \tag{10-7}$$

Using the fact that the noise pairs $(w_r, w_\alpha)$, $(w_{\dot{r}}, w_\alpha)$, and $(w_{\dot{c}}, w_\alpha)$, are each statistically independent (by assumption; see equation (10-5)), and using the identities

$$E_{w_\alpha}\{\cos(\alpha + w_\alpha)\} = e^{-\sigma_\alpha^2/2}\cos(\alpha), \tag{10-8}$$

$$E_{w_\alpha}\{\sin(\alpha + w_\alpha)\} = e^{-\sigma_\alpha^2/2}\sin(\alpha), \tag{10-9}$$

it is straightforward to show

$$E_{\tilde{\mathbf{z}}_z}\{g(\mathbf{z}_z)\} = e^{-\sigma_\alpha^2/2}\mathbf{x}_x. \tag{10-10}$$

Substituting this result into equation (6-12) yields

$$\mathbf{B}\,e^{-\sigma_\alpha^2/2}\,\mathbf{x}_\chi = \mathbf{x}_\chi. \tag{10-11}$$

From this result, it is apparent the debiasing matrix, B, is given by

$$\mathbf{B} = e^{\sigma_\alpha^2/2}\,\mathbf{I}_{4\times 4}. \tag{10-12}$$

Hence, the debiased, converted measurement for this example is given by

$$\mathbf{z}_\chi = \mathbf{B} g(\mathbf{z}_z) = e^{\sigma_\alpha^2/2} \begin{bmatrix} z_r \cos z_\alpha \\ z_r \sin z_\alpha \\ z_{\dot r} \cos z_\alpha - z_{\dot c} \sin z_\alpha \\ z_{\dot r} \sin z_\alpha + z_{\dot c} \cos z_\alpha \end{bmatrix}. \tag{10-13}$$

By assumption 3, any of the raw measurements in this expression that are not observed can be replaced by predicted measurements, based on the predicted state estimate, $\hat{\mathbf{x}}_\chi(k|k-1)$.

## 10.3 CONVERTED MEASUREMENT PRECISION MATRIX

For this example, the converted measurement precision matrix, $\bar{\mathbf{R}}_\chi^{-1}$, is obtained using the procedure developed in section 7, starting with determining the covariance matrix, $\hat{\mathbf{R}}_\chi$, given by equation (7-9), which implicitly assumes each polar coordinate—range, bearing, range rate, and cross-range rate—are observed. For this example, a useful approximation to $\hat{\mathbf{R}}_\chi$ can be derived analytically (see reference 9 or chapter 4 of reference 7). The derivation is straightforward, but algebraically tedious, and is not included here. For the simulations presented in section 11, these covariance matrices are instead evaluated using the numerical methods discussed in section 8.

Once the covariance matrix, $\hat{\mathbf{R}}_\chi$, is obtained, the converted measurement precision matrix, $\bar{\mathbf{R}}_\chi^{-1}$, to be used in the PKF recursions presented in section 5, is evaluated using equation (7-11) for the appropriate form of the information zeroing matrix, $W$. For instance, if only range and bearing measurements are observed, then the upper left block of $W$ is set to $\mathbf{I}_{2\times 2}$, with all other elements set to zero. Likewise, if Doppler information in the form of range rate measurements are also observed, then the upper left block of $W$ is set to $\mathbf{I}_{3\times 3}$, with all other elements set to zero.

The converted measurement precision matrix, $\bar{\mathbf{R}}_\chi^{-1}$, as given by equation (7-11), requires evaluation of the inverse Jacobian matrices, $\mathbf{J}_h^{-1}$ and $\mathbf{J}_g^{-1}$, which, by assumption 2, are equivalent to $\mathbf{J}_g$ and $\mathbf{J}_h$, respectively. For this problem, $\mathbf{J}_g$ and $\mathbf{J}_g^{-1}$ are the easiest to compute. From equation (10-2), the Jacobian of $g$, evaluated at the predicted state in the measurement space, i.e., $\hat{\mathbf{x}}_z = \begin{bmatrix} \hat r & \hat\alpha & \hat{\dot r} & \hat{\dot c} \end{bmatrix}^t$, is given by

$$\mathbf{J}_g(\hat{\mathbf{x}}_z) = \begin{bmatrix} \cos\hat\alpha & -\hat r \sin\hat\alpha & 0 & 0 \\ \sin\hat\alpha & \hat r \cos\hat\alpha & 0 & 0 \\ 0 & -\hat{\dot r}\sin\hat\alpha - \hat{\dot c}\cos\hat\alpha & \cos\hat\alpha & -\sin\hat\alpha \\ 0 & \hat{\dot r}\cos\hat\alpha - \hat{\dot c}\sin\hat\alpha & \sin\hat\alpha & \cos\hat\alpha \end{bmatrix}, \tag{10-14}$$

and its inverse by

$$\mathbf{J}_g^{-1}(\hat{\mathbf{x}}_z) = \begin{bmatrix} \cos\hat{\alpha} & \sin\hat{\alpha} & 0 & 0 \\ -\hat{r}^{-1}\sin\hat{\alpha} & \hat{r}^{-1}\cos\hat{\alpha} & 0 & 0 \\ -\hat{\dot{c}}\hat{r}^{-1}\sin\hat{\alpha} & \hat{\dot{c}}\hat{r}^{-1}\cos\hat{\alpha} & \cos\hat{\alpha} & \sin\hat{\alpha} \\ \hat{\dot{r}}\hat{r}^{-1}\sin\hat{\alpha} & -\hat{\dot{r}}\hat{r}^{-1}\cos\hat{\alpha} & -\sin\hat{\alpha} & \cos\hat{\alpha} \end{bmatrix}. \quad (10\text{-}15)$$

In the range/bearing case, it is worth noting that partitioning $\mathbf{J}_g^{-t}(\hat{\mathbf{x}}_z)$ as in equation (7-13) yields the block matrices

$$\mathbf{J}_{g,m}^{-t}(\hat{\mathbf{x}}_z) = \begin{bmatrix} \cos\hat{\alpha} & -\hat{r}^{-1}\sin\hat{\alpha} \\ \sin\hat{\alpha} & \hat{r}^{-1}\cos\hat{\alpha} \end{bmatrix}, \quad (10\text{-}16)$$

$$\mathbf{J}_{g,um}^{-t}(\hat{\mathbf{x}}_z) = \begin{bmatrix} 0 & 0 \\ 0 & 0 \end{bmatrix}, \quad (10\text{-}17)$$

which, in turn, when substituted into equation (7-13), yields the converted measurement precision matrix

$$\bar{\mathbf{R}}_x^{-1} = \begin{bmatrix} \mathbf{J}_{g,m}^{-t}(\hat{\mathbf{x}}_z)\,\bar{\mathbf{R}}_{z_m}^{-1}\,\mathbf{J}_{g,m}^{-1}(\hat{\mathbf{x}}_z) & 0 & 0 \\ & 0 & 0 \\ 0 & 0 & 0 & 0 \\ 0 & 0 & 0 & 0 \end{bmatrix}. \quad (10\text{-}18)$$

Hence, in this case, the PKF state update equation, equation (5-5), will not be a function of the non-informative measurements, $z_{\dot{r}}$ and $z_{\dot{c}}$, since pre-multiplying the measurement residual vector in equation (5-5) by $\bar{\mathbf{R}}_x^{-1}$ zeros their contributions to the update. The same is not true in the range/bearing/Doppler case, where partitioning $\mathbf{J}_g^{-t}(\hat{\mathbf{x}}_z)$ as in equation (7-13) yields the block matrices

$$\mathbf{J}_{g,m}^{-t}(\hat{\mathbf{x}}_z) = \begin{bmatrix} \cos\hat{\alpha} & -\hat{r}^{-1}\sin\hat{\alpha} & -\hat{\dot{c}}\hat{r}^{-1}\sin\hat{\alpha} \\ \sin\hat{\alpha} & \hat{r}^{-1}\cos\hat{\alpha} & \hat{\dot{c}}\hat{r}^{-1}\cos\hat{\alpha} \\ 0 & 0 & \cos\hat{\alpha} \end{bmatrix}, \quad (10\text{-}19)$$

$$\mathbf{J}_{g,um}^{-t}(\hat{\mathbf{x}}_z) = \begin{bmatrix} 0 & 0 & \sin\hat{\alpha} \end{bmatrix}, \quad (10\text{-}20)$$

which, in turn, yield a converted measurement precision matrix, $\bar{\mathbf{R}}_x^{-1}$, which, while not full rank, is, in general, fully populated. Thus, in this case, the PKF state update equation, equation (5-5), will be a function of the non-informative measurement of cross-range rate, $z_{\dot{c}}$.

## 11. SIMULATIONS

Performance comparisons of the PKF and standard implementations of the UKF and EKF for the polar-to-Cartesian coordinate conversion problem are presented in this section for two cases. In case one, a stationary observer measures range and bearing to a constant velocity target at a fixed update rate. In case two, the observer measures the target's range rate in addition to its range and bearing. For each case, three Monte Carlo experiments are performed, with 1000 trials for each experiment. The outputs of the six experiments are shown in figures 1 through 6 on separate pages at the end of this section.

For each experiment, each trial consists of 100 measurements at a 2-second update rate. In both cases, the observed range and bearing measurements are assumed to have standard deviations, $\sigma_r = 30$ meters and $\sigma_\alpha = 0.0873$ radians (5 degrees), respectively. For case one, the unobserved range rate and cross-range rate measurements are assumed to have standard deviations, $\sigma_{\dot{r}} = \sigma_{\dot{c}} = 10$ meters/second. For case two, the measured range rate is assumed to have standard deviation, $\sigma_{\dot{r}} = 0.1$ meters/second. In both cases, a range and range rate correlation coefficient, $\rho = -0.2$, is assumed.

For each trial of each Monte Carlo experiment, the true target state in Cartesian coordinates at time index, $k = 0$, that is, $\mathbf{x}_x(0)$, is randomly sampled from the following distributions:

• initial range to the target is normally distributed with mean 4000 meters and standard deviation 30 meters;

• initial bearing to the target is uniformly distributed between 0 and $2\pi$;

• target heading is uniformly distributed between 0 and $2\pi$;

• target speed is chi-square distributed with 2 degrees of freedom, and scaled by a factor of 10 meters/second.

Accordingly, the initial target state covariance matrix in Cartesian coordinates for each filter in each trial is taken as

$$\mathbf{P}_x(0|0) = \begin{bmatrix} 30^2 & 0 & 0 & 0 \\ 0 & 30^2 & 0 & 0 \\ 0 & 0 & 10^2 & 0 \\ 0 & 0 & 0 & 10^2 \end{bmatrix}, \qquad (11\text{-}1)$$

and the corresponding initial target state estimate, $\hat{\mathbf{x}}_x(0|0)$, is drawn from a multivariate normal distribution with mean, $\mathbf{x}_x(0)$, and covariance matrix, $\mathbf{P}_x(0|0)$.

For a fixed, 2-second update rate, the $4 \times 4$ state transition matrix, $A(k)$, for this constant velocity target tracking problem is given by the constant matrix,

$$A(k) = \begin{bmatrix} 1 & 0 & 2 & 0 \\ 0 & 1 & 0 & 2 \\ 0 & 0 & 1 & 0 \\ 0 & 0 & 0 & 1 \end{bmatrix}. \qquad (11\text{-}2)$$

Assuming a white noise acceleration model for the process noise term, $\tilde{\mathbf{q}}_x(k)$, in equation (3-1), and the discretized form of the process noise covariance matrix, $Q_x(k)$, in equation (5-2) (see section 6.2.2 of reference 1), this matrix is also constant and is taken as

$$Q_x(k) = 2\,(0.44)^2 \begin{bmatrix} \frac{4}{3} & 1 \\ 1 & 1 \end{bmatrix}, \qquad (11\text{-}3)$$

where the process noise intensity is taken as $0.44^2$.

The PKF implementation uses a multiplicative debiasing function. The debiasing matrix, B, from equation (10-12) is given by

$$\mathbf{B} = e^{0.0873^2/2} \mathbf{I}_{4\times 4}, \tag{11-4}$$

where the bearing measurement standard deviation $\sigma_\alpha = 0.0873$ radians.

Finally, the implementations of the UKF and the PKF use the 5th order McNamee-Stenger rule from reference 24, as described in reference 20, to generate the sigma points and corresponding weights for the SPTs required by each filter.

Figures 1 through 6 at the end of this section show the results of the six Monte Carlo experiments for this example. The first subplot in each figure shows the average normalized estimation error squared (ANEES) (further normalized by the state dimension, $N$) for each state estimate (for each filter) at each update, $k$, defined here as

$$\psi(k) = \frac{1}{NL} \sum_{l=1}^{L} \left[\hat{\mathbf{x}}_x^{(l)}(k|k) - \mathbf{x}_x(k)\right]^t \left[\mathbf{P}_x^{(l)}(k|k)\right]^{-1} \left[\hat{\mathbf{x}}_x^{(l)}(k|k) - \mathbf{x}_x(k)\right], \tag{11-5}$$

where $L$ denotes the number of Monte Carlo trials (1000 for each experiment), and the superscript, $(l)$, denotes the filter output for trial, $l$. The ANEES is a measure of filter consistency or, in words, how well the filter's estimate of its error matches the actual estimation error, on average. For the standard (linear Gaussian) Kalman filter, the statistic $NL\psi(k)$ is chi-square distributed with $NL$ degrees of freedom, with mean equal to $NL$ and variance equal to $2NL$. Thus, for the standard Kalman filter, the expected value and variance of $\psi(k)$ equal 1 and $2/(NL)$, respectively.

The second and third subplots in each figure show the MSE in the position and velocity estimates, respectively, for each filter at each update, $k$. Before defining the MSE, let $\mathbf{H}_\mathcal{S}$ denote the matrix formed by the the consecutive rows of the $N \times N$ identity matrix corresponding to the row indices in the ordered set, $\mathcal{S} \subseteq \{1, 2, \ldots, N\}$. Choosing $\mathcal{S} = \{1, 2\}$ yields

$$\mathbf{H}_{\{1,2\}} = \begin{bmatrix} 1 & 0 & 0 & 0 \\ 0 & 1 & 0 & 0 \end{bmatrix}. \tag{11-6}$$

As a result, $\mathbf{H}_\mathcal{S}\mathbf{x}_x(k)$ selects only the position elements in $\mathbf{x}_x(k)$. Similarly, setting $\mathcal{S} = \{3, 4\}$, yields an $\mathbf{H}_\mathcal{S}$ that selects the velocity components in $\mathbf{x}_x(k)$. Thus, for any given $\mathcal{S}$, the mean squared error is defined as

$$\eta_\mathcal{S}(k) = \frac{1}{L} \sum_{l=1}^{L} \left[\hat{\mathbf{x}}_x^{(l)}(k|k) - \mathbf{x}_x(k)\right]^t \mathbf{H}_\mathcal{S}^t \mathbf{H}_\mathcal{S} \left[\hat{\mathbf{x}}_x^{(l)}(k|k) - \mathbf{x}_x(k)\right], \tag{11-7}$$

The second and third subplots in each figure show the MSE in the position and velocity estimates, corresponding to $\mathcal{S} = \{1, 2\}$ and $\mathcal{S} = \{3, 4\}$, respectively. For reference, each of these plots also include the corresponding error derived from the posterior Cramér-Rao lower bound (PCRLB) for the state coordinates of interest; ideally, the MSE of an efficient estimator would meet this lower bound from above. For each experiment, the PCRLB is taken as the PCRLB averaged over all trials, as it is evaluated at the true target trajectory, and the true initial target state is randomized over the trials.

The 95% confidence intervals for the ANEES and MSE are computed for each filter and experiment, and are shown in each subplot of each figure as dashed lines. The 95% confidence intervals for the ANEES are plotted as black dashed lines in the first subplot of each figure. The 95% confidence intervals for the MSE of each filter are plotted as color-matched dashed lines in the second and third subplots of each figure. The 95% confidence intervals for the MSE for each filter are computed using the number of runs where the filter did not lose track. The track loss numbers for each filter and experiment are listed in table 1.

Figures 1 through 3 compare the performance of the PKF, UKF, and EKF for the case of a stationary observer measuring range and bearing to a constant velocity target for three Monte Carlo experiments with otherwise identical parameters. In this case, both the PKF and UKF vastly outperform the EKF, and the PKF outperforms the UKF, on average, often by a considerable margin. In particular, the PKF is remarkably consistent at all updates and across all three experiments, as its ANEES values are largely contained within the 95% confidence interval. The PKF in this case is also remarkably efficient because the 95% confidence intervals about its MSE values largely contain the PCRLB at each update across all three experiments.

Likewise, figures 4 through 6 compare the performance of the PKF, UKF, and EKF for the case of a stationary observer measuring range, bearing, and range rate to a constant velocity target for three Monte Carlo experiments with otherwise identical parameters. In this case, the PKF vastly outperforms the EKF and has better performance than the UKF. In particular, the PKF produces ANEES values below the lower 95% confidence interval, with values closer to 0.9 than 1, on average. Thus, in this case, the PKF is slightly pessimistic with respect to its self-assessment of estimation error (i.e., the state covariance matrices are a little too large), which means it is less likely to lose track. In contrast, the UKF produces ANEES values above the 95% confidence interval, which means the UKF is optimistic; consequently, it loses some tracks. The PKF in this case is also reasonably efficient: the 95% confidence intervals for the PKF MSE values largely contain the PCRLB at each update across all three experiments. This is not true for the UKF's 95% confidence interval on its position estimates. As hoped, the addition of range-rate to the measurement set reduces the MSE errors in the PKF estimates. However, the opposite is true for the EKF, indicating the EKF does a worse job of accounting for the nonlinearity in the mapping of range rate to the Cartesian target state. The performance of the UKF improves with the addition of the range rate too, but its performance is not as good as the PKF. In particular, the fact that the ANEES values for the UKF are all above the 95% confidence interval suggest it too does not account for the nonlinear mapping as well as the PKF.

Finally, table 1 compares the performance of the three filters for each of the six experiments in terms of the number of lost tracks for each filter in each experiment. Remarkably, the PKF does not lose track in any of the experiments. For the UKF and EKF, table 1 shows the 95% confidence intervals (CIs) on the number of lost tracks (it is not possible to calculate the 95% confidence intervals for the PKF because the PKF did not lose track). The UKF has a reasonable track loss rate of less than 3% in all six experiments. While the EKF also does reasonable well in this respect for the range/bearing case, with a track loss rate of less than 7%, it does poorly with the addition of range rate measurements, where it exhibits a track loss rate of around 50%.

*Table 1. Numbers of Lost Track (out of 1000) for Each Filter and Monte Carlo Experiment*

| Measurements | Experiment | PKF | UKF (95% CI) | EKF (95% CI) |
|---|---|---|---|---|
| Range, Bearing | 1 | 0 | [12, 30] | [34, 60] |
|  | 2 | 0 | [11, 29] | [38, 66] |
|  | 3 | 0 | [8, 24] | [22, 44] |
| Range, Bearing, Range Rate | 4 | 0 | [17, 37] | [462, 524] |
|  | 5 | 0 | [11, 27] | [471, 533] |
|  | 6 | 0 | [11, 29] | [497, 559] |

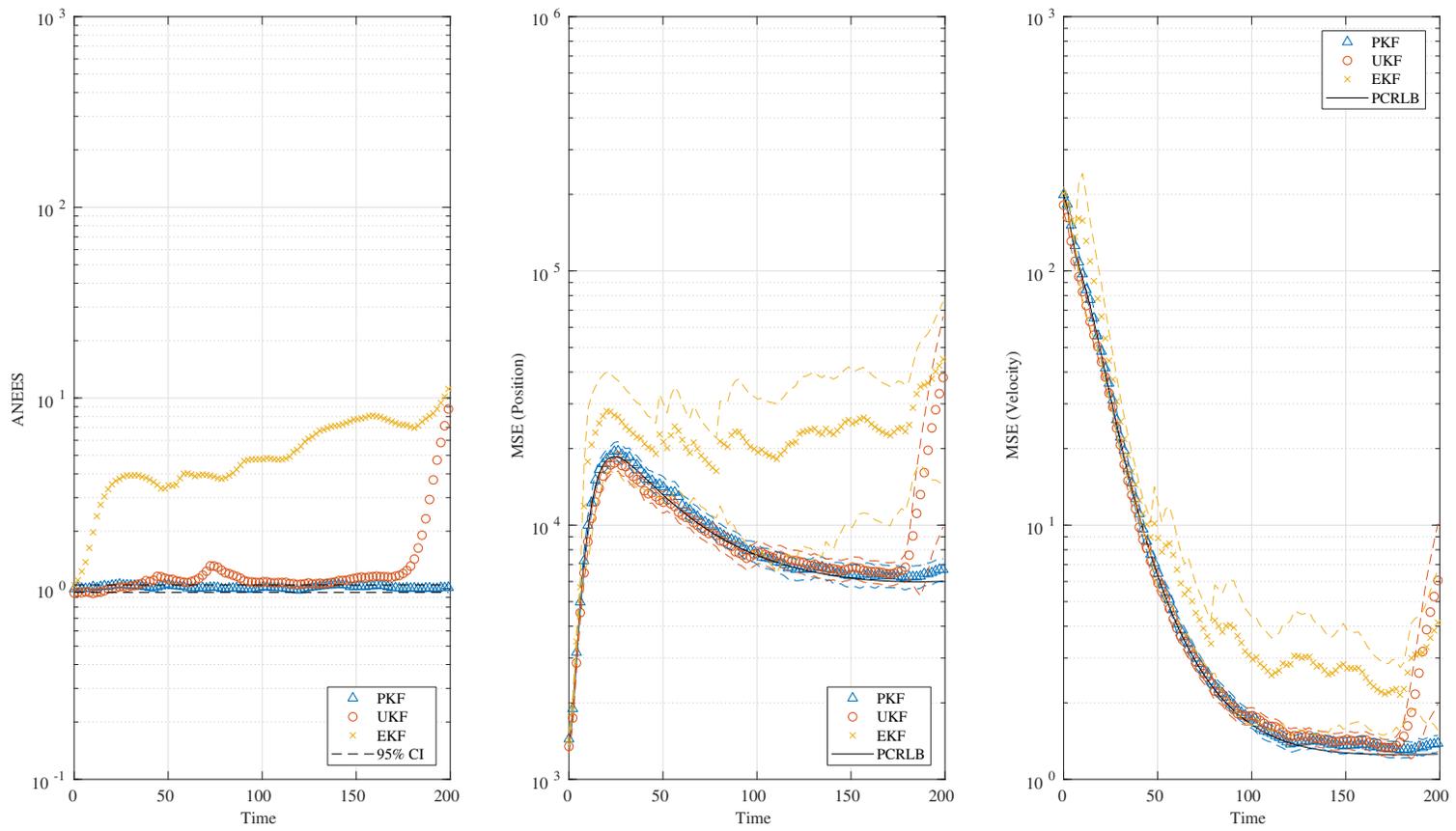

*Figure 1. Filter Performance Comparison with Range and Bearing Measurements (Monte Carlo Experiment 1)*

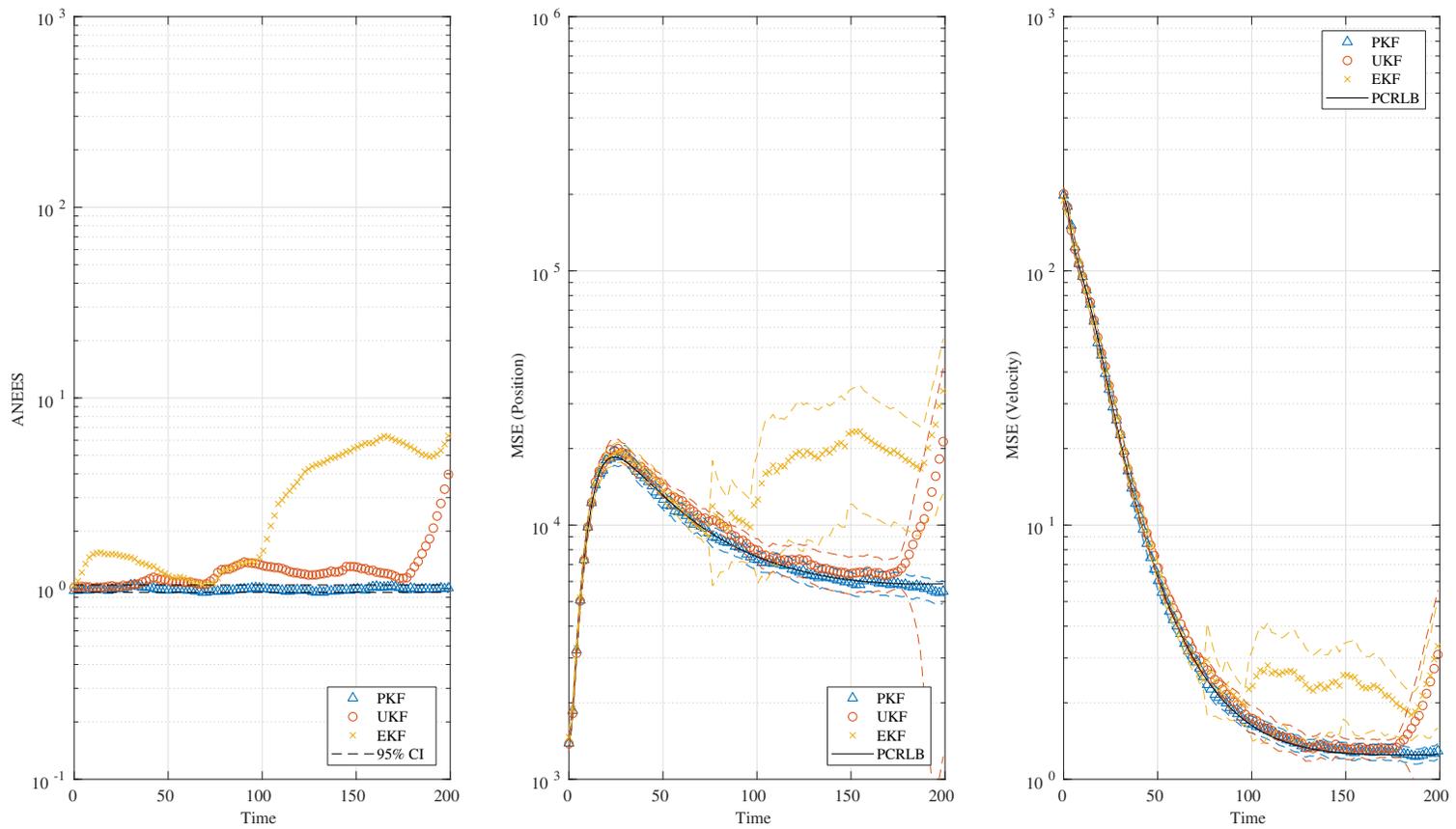

*Figure 2. Filter Performance Comparison with Range and Bearing Measurements (Monte Carlo Experiment 2)*

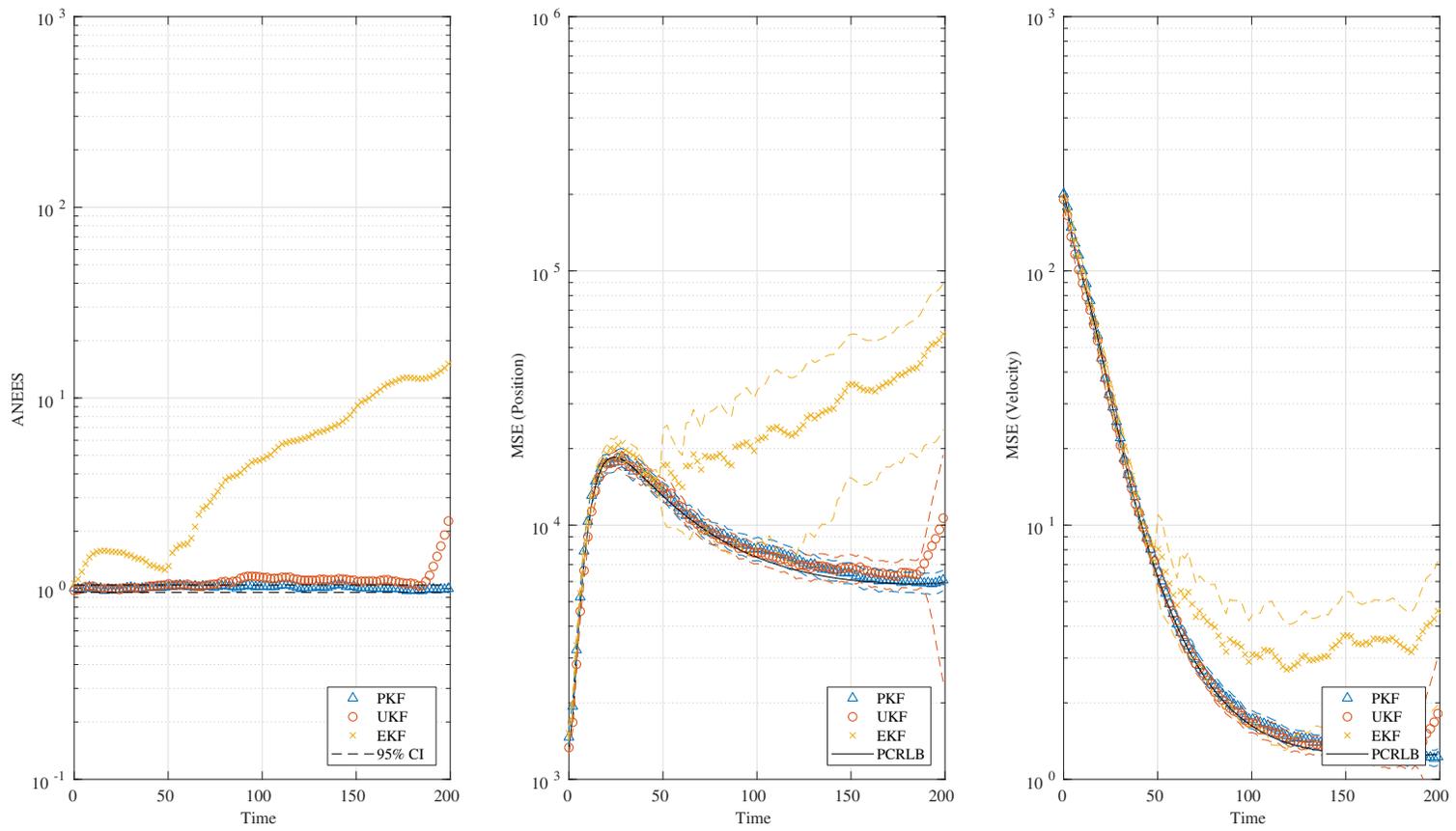

*Figure 3. Filter Performance Comparison with Range and Bearing Measurements (Monte Carlo Experiment 3)*

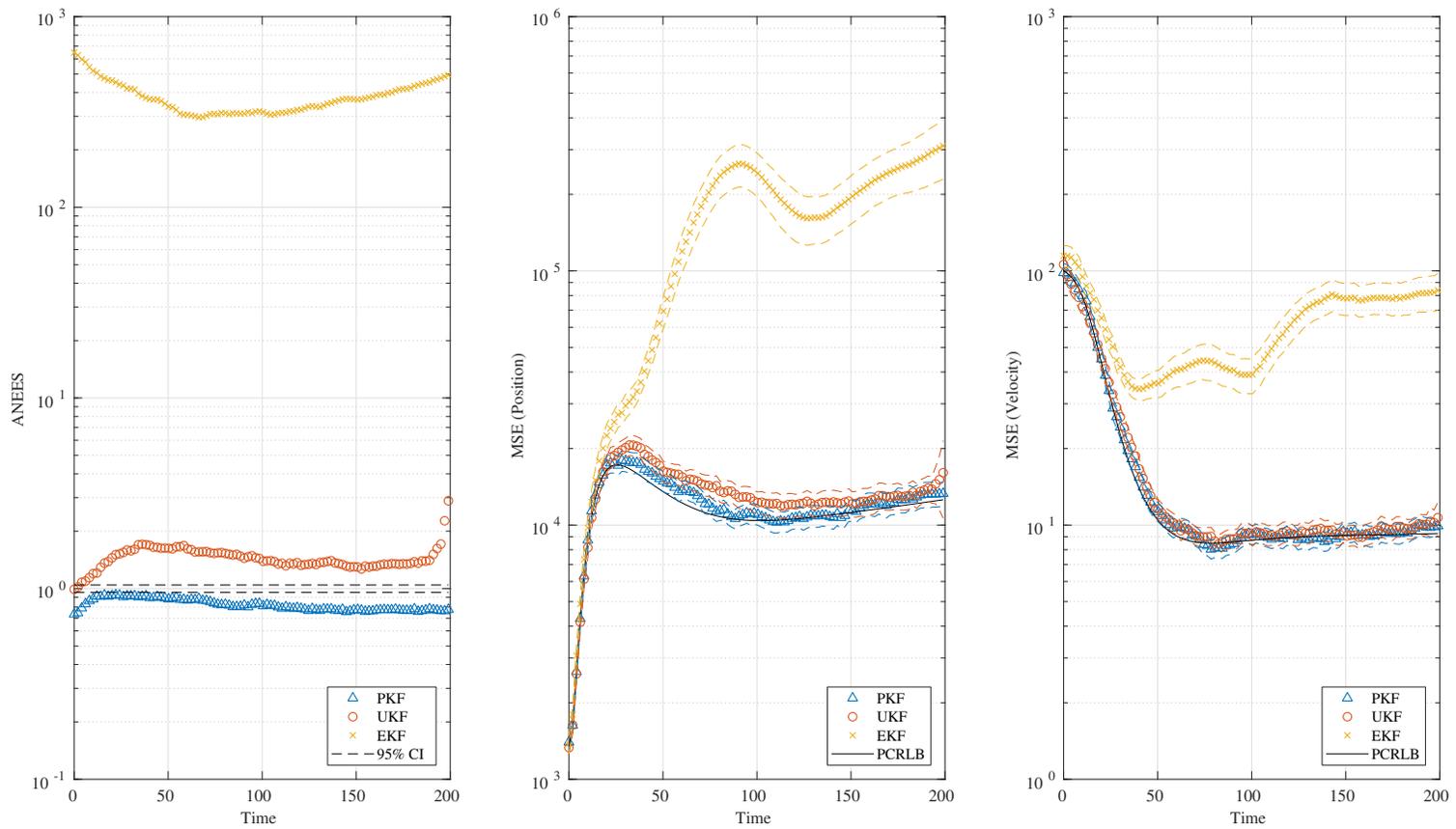

*Figure 4. Filter Performance Comparison with Range, Bearing, and Range-Rate Measurements (Monte Carlo Experiment 4)*

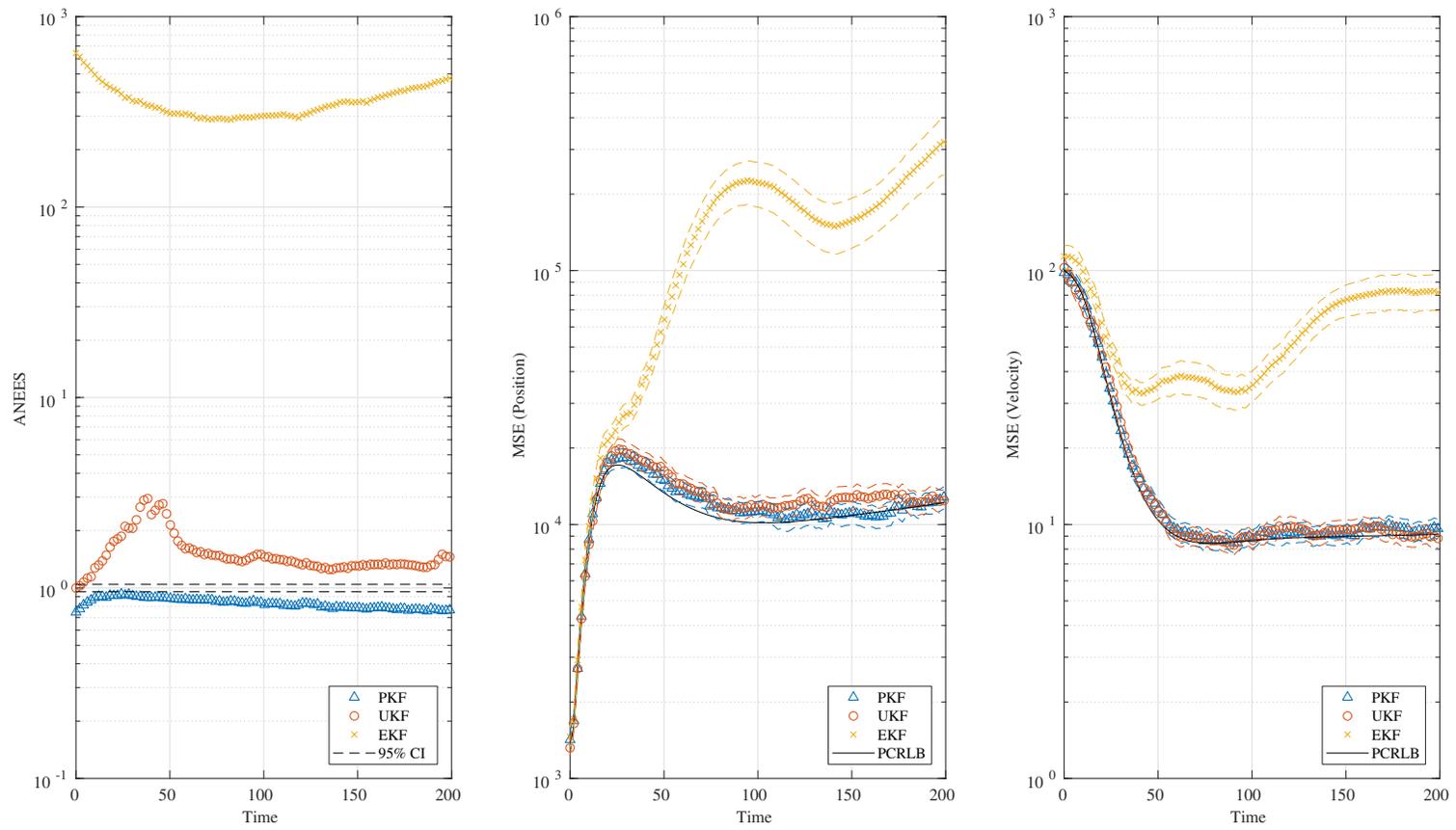

*Figure 5. Filter Performance Comparison with Range, Bearing, and Range-Rate Measurements (Monte Carlo Experiment 5)*

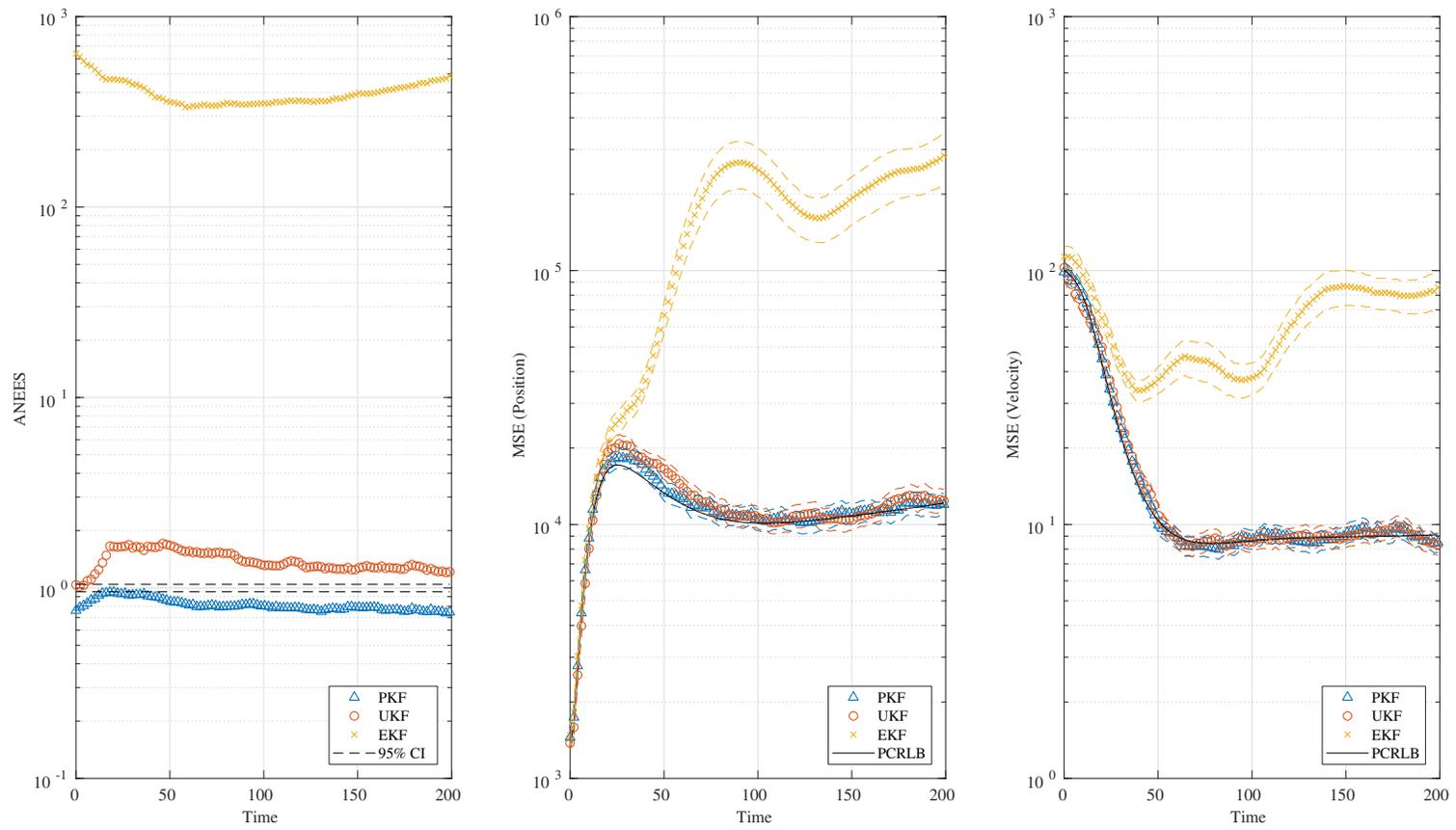

*Figure 6. Filter Performance Comparison with Range, Bearing, and Range-Rate Measurements (Monte Carlo Experiment 6)*

## 12. SUMMARY AND CONCLUSION

In this report, the measurement conversion procedure developed in reference 7 and the references cited therein is generalized to the class of Kalman filtering problems with a linear state equation and nonlinear measurement equation for which a bijective mapping exists between the state and measurement coordinate systems. Many practical target tracking problems are among this class, including those that feature measurements in polar coordinates and states in Cartesian coordinates. The generalized measurement conversion procedure leads to a version of the information form of the Kalman filter, with recursions that are explicit functions of the converted measurement precision matrix. Hence, this generalized converted measurement Kalman filter is dubbed the precision Kalman filter, or PKF.

This precision matrix formulation of the converted measurement Kalman filter allows for the explicit use of noninformative measurements—that is, measurements with infinite variances—in the measurement conversion equations. This is achieved by setting their corresponding entries in the converted measurement precision matrix (mapped to the measurement coordinate system) to zeros. While this zeroing operation prevents the noninformative measurements from directly adding information, they may influence off-diagonal terms in the converted measurement precision matrix, as demonstrated by introduction of the non-informative cross-range rate measurements in the range, bearing, and range rate example. Expectations required by the measurement conversion process that cannot be solved analytically can be readily computed using numerical integration (e.g., the sigma point transform) methods.

For the common polar-to-Cartesian measurement conversion problem of tracking a constant velocity target from range and bearing, or range, bearing, and range-rate measurements, the PKF is shown in simulations to outperform both the standard UKF and EKF formulations in terms of filter consistency, efficiency, and track loss. In fact, the PKF exhibits no track loss, ANEES values consistently near unity, and MSEs in position and velocity that closely track their PCRLB values.

The PKF is widely applicable to a variety of practical filtering problems involving nonlinear measurement equations that satisfy the requisite assumptions, including the bistatic tracking problem and the inertial navigation problem. Future work will focus on these two applications in particular.